\documentclass[twocolumn,english,amsmath,amssymb,superscriptaddress,nofootinbib]{revtex4-1}
\usepackage[T1]{fontenc}
\usepackage[latin9]{inputenc}
\usepackage{amsmath}
\usepackage{amssymb}
\usepackage{graphicx}

\usepackage{color}

\usepackage{ulem}

\makeatletter

\providecommand{\tabularnewline}{\\}

\makeatother

\usepackage{babel}
\begin{document}
\title{
Cavity Quantum Electrodynamics Effects of Optically Cooled Nitrogen-Vacancy Centers Coupled to a High Frequency Microwave Resonator
}
\author{Yuan Zhang}
\email{yzhuaudipc@zzu.edu.cn}
\address{Henan Key Laboratory of Diamond Optoelectronic Materials and Devices, Key Laboratory of Material Physics Ministry of Education, School of Physics and Microelectronics, Zhengzhou University, Daxue Road 75, Zhengzhou 450052 China }
\author{Qilong Wu}
\address{Henan Key Laboratory of Diamond Optoelectronic Materials and Devices, Key Laboratory of Material Physics Ministry of Education, School of Physics and Microelectronics, Zhengzhou University, Daxue Road 75, Zhengzhou 450052 China }
\author{Hao Wu}
\address{Center for Quantum Technology Research and Key Laboratory of Advanced Optoelectronic Quantum Architecture and Measurements (MOE), School of Physics, Beijing Institute of Technology, Beijing 100081, China and Beijing Academy of Quantum Information Sciences, Beijing 100193, China}
\author{Xun Yang}
\address{Henan Key Laboratory of Diamond Optoelectronic Materials and Devices, Key Laboratory of Material Physics Ministry of Education, School of Physics and Microelectronics, Zhengzhou University, Daxue Road 75, Zhengzhou 450052 China }
\author{Shi-Lei Su}
\address{Henan Key Laboratory of Diamond Optoelectronic Materials and Devices, Key Laboratory of Material Physics Ministry of Education, School of Physics and Microelectronics, Zhengzhou University, Daxue Road 75, Zhengzhou 450052 China }
\author{Chongxin Shan}
\email{cxshan@zzu.edu.cn}
\address{Henan Key Laboratory of Diamond Optoelectronic Materials and Devices, Key Laboratory of Material Physics Ministry of Education, School of Physics and Microelectronics, Zhengzhou University, Daxue Road 75, Zhengzhou 450052 China }
\author{Klaus M{\o}lmer}
\email{moelmer@phys.au.dk}
\address{Center for Complex Quantum Systems, Department of Physics and Astronomy,
Aarhus University, Ny Munkegade 120, DK-8000 Aarhus C, Denmark and Aarhus Institute of Advanced Studies, Aarhus University, H{\o}egh-Guldbergs Gade 6B, DK-8000 Aarhus C, Denmark}
\begin{abstract}
Recent experiments demonstrated the cooling of a microwave mode of a high-quality dielectric resonator coupled to optically cooled nitrogen-vacancy (NV) spins in diamond. Our recent theoretical study [arXiv:2110.10950] pointed out the cooled NV spins can be used to realize cavity quantum electrodynamics effects (C-QED) at room temperature. In this article, we propose to modify the setup used in a recent diamond maser experiment [Nature \textbf{55}, 493-496 (2018)], which features a higher spin transition frequency, a lower spin-dephasing rate and a stronger NV spins-resonator coupling, to realize better microwave mode cooling and the room-temperature CQED effects. To describe more precisely the optical spin cooling and the collective spin-resonator coupling, we extend the standard Jaynes-Cumming model to account for the rich electronic and spin levels of the NV centers. Our calculations show that for the proposed setup it is possible to cool the microwave mode from $293$ K (room temperature) to $116$ K, which is about $72$ K lower than the previous records, and to study the intriguing dynamics of the CQED effects under the weak-to-strong coupling transition by varying the laser power.  With simple modifications, our model can be applied to, e.g., other solid-state spins or triplet spins of pentacene molecules, and to investigate other effects, such as the operations of pulsed and continuous-wave masing. 
\end{abstract}
\maketitle

\section{Introduction}


Nitrogen vacancy (NV) centers in diamond constitute a prototypical solid-state spin system~\citep{Eisenach}, which offers many applications, such as sensing of magnetic fields~\citep{BarryJF,Shi}, electric fields~\citep{Doherty}, local strain~\citep{Barson}, temperature~\citep{Kucsko}, and quantum information processing and computation~\citep{Tao,Ladd}. These applications benefit from  unique properties of the NV centers, namely, a spin-1 triplet electronic ground state with long coherence time at room temperature, and easy initialization and readout of the spin state by optical means.

\begin{figure}
\begin{centering}
\includegraphics[scale=0.37]{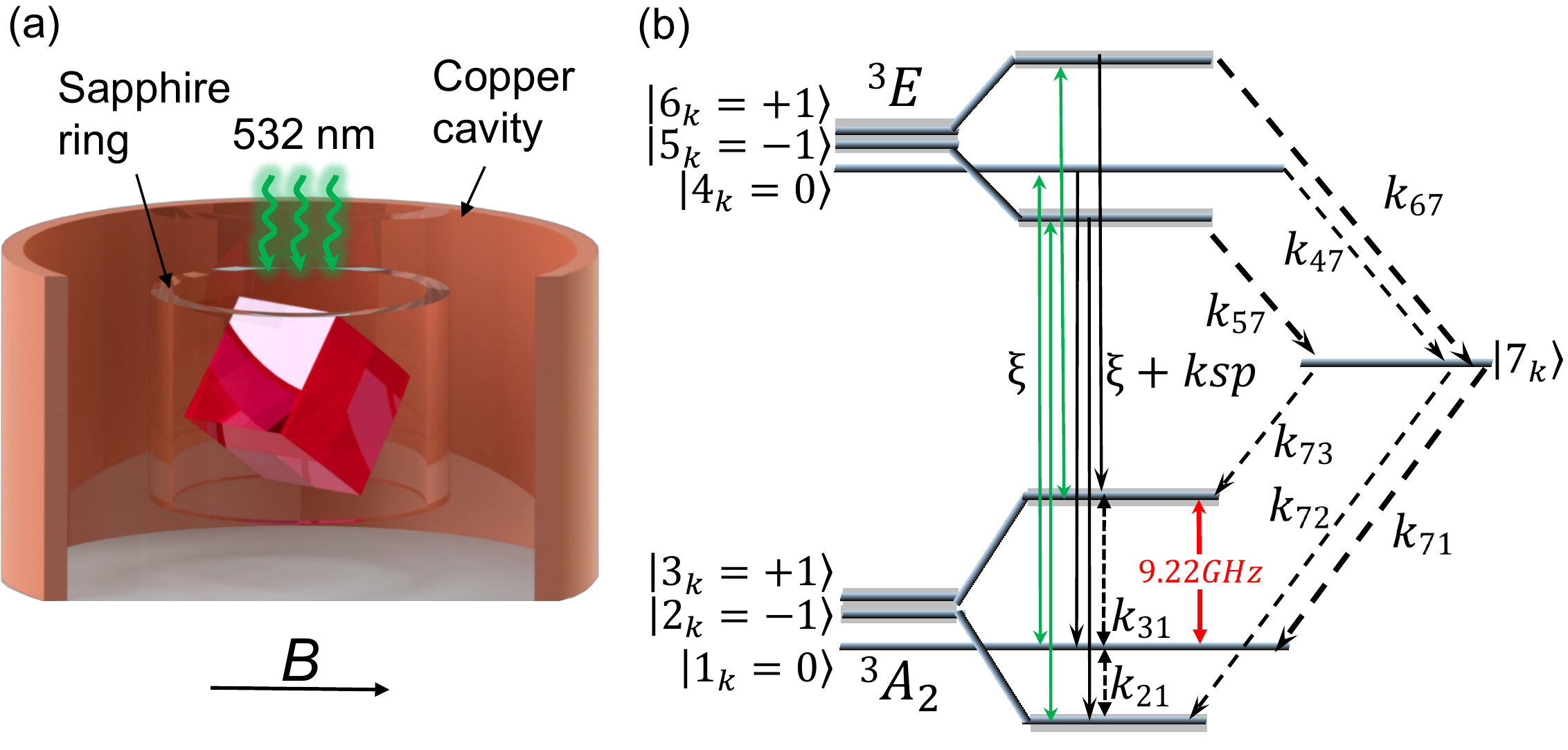}
\par\end{centering}
\caption{\label{fig:sys} Proposed system and energy diagram. Panel (a) shows the setup modified from the diamond maser experiment ~\citep{Breeze}  with nitrogen-vacancy (NV) centers in diamond excited by a 532 nm laser and coupled to a single-crystal sapphire dielectric microwave ring resonator inside a copper cylindrical cavity in the presence of a magnetic field. Panel (b) shows the involved processes for multiple energy levels of the NV centers, such as the downward and upward Zeeman shift of the spin level with projections $-1$ and $+1$, the optical excitation and stimulated emission with rate $\xi$ (green double-headed arrows), the spontaneous emission with rate $k_{sp}$ (downward black arrows), the inter-system crossing with rates $k_{i7},k_{7j}$ (with $i=4,5,6$ and $j=1,2,3$, tilted dashed arrows), the spin-lattice relaxation with rates   $k_{31}\approx k_{13}, k_{21} \approx k_{12}$ (double-head dashed arrows), and the spin-dephasing rate $\chi_2,\chi_3$ (shaded gray background). In the  experiment~\citep{Breeze}, the population inversion between the $0$ and $-1$ spin level is explored for the maser operation, but, here, we explore the negative inversion between the $+1$ and $0$ spin level to cool the microwave resonator with frequency $2\pi\times9.22$ GHz (red solid arrow). The values of various rates are specified in Tab. \ref{tab:param} in the Appendix \ref{sec:params}, and more details are provided in the main text. }
\end{figure}

Besides spin initialization, the NV spin ensemble can also be cooled with the optical pumping, and the cooled spin ensemble was recently used to cool a resonantly coupled microwave mode of a $2.87$ GHz dielectric resonator from $293$ K (room temperature) down to $188$ K ~\citep{NgW}. In contrast to conventional cooling with bulky dilution refrigerators, this cooling mechanism may bring significant technical advances as it can be achieved with a bench top device in a room temperature laboratory environment. A sufficiently cooled microwave mode may approach pure quantum states and permit study of  quantum entanglement~\citep{SHaroche}, quantum gate operations~\citep{Henschel} and quantum thermodynamics~\citep{Klatzow}, and it can also improve the measurement sensitivity in electron and nuclear spin resonance experiments~\citep{Shi,Staudacher}. Encouraged by the unambiguous demonstration of cavity quantum electrodynamics (C-QED) effects with the triplet spins of pentacene molecules at room temperature~\citep{JDBreeze}, we showed in a recent publication~\citep{zhangyuan1}, that the optically cooled NV spins permit the realization of C-QED effects at room temperature, such as Rabi oscillations, Rabi splittings and stimulated superradiance.




So far, the microwave mode cooling by NV center spins is limited by the weak spin-microwave mode coupling compared with the large spin-dephasing rate and the high ambient excitation of the low frequency microwave mode. To improve the microwave mode cooling and explore the CQED effects, in this article, we propose to study the setup shown in Fig. \ref{fig:sys}, where a magnetic field  Zeeman-splits the NV spin levels, and the $+1 \to 0$ spin transition couples resonantly to a dielectric microwave resonator with higher frequency $9.22$ GHz (Fig. \ref{fig:sys}). In such a setup, the NV spin ensemble has a lower spin-dephasing rate, the resonator microwave mode has a lower initial thermal excitation, and their coupling is also much stronger, which can work together  to achieve a much lower mode temperature, and the room-temperature CQED effects. Note that the proposed setup is different from the one in the diamond maser experiment~\citep{Breeze}, which utilizes the $0 \to -1$ spin transition. 


To describe more precisely the optical spin cooling and the NV spin-resonator  collective coupling, we extend the standard Jaynes-Cumming (JC) model for many two-level systems as used in Ref.~\citep{zhangyuan1} to account for all the electronic and spin levels of NV centers in Sec. \ref{sec:multilevelJC}, see Fig. \ref{fig:sys} (b). To approach the experiments as close as possible, we consider the systems with trillions of NV centers by solving the extended model with mean-field approach~\citep{KDebnath2018,YZhang2021,QWu2021}. To treat properly the collective effects, we consider the mean-field quantities up to second order, which include the spin-photon and spin-spin quantum correlations. 

To determine the conditions to achieve better microwave mode cooling,  we investigate the influence of the laser excitation on the evolution of the ground state spin-level populations and the microwave mode excitation in Sec. \ref{sec:modecooling}. Our calculations predict the cooling of the microwave mode from $293$ K (room temperature) to $116$ K, about $72$ K below the record achieved so far with NV centers in diamond \citep{NgW}. A simpler rate equation treatment yields a temperature of $87$ K, and we explain the difference of the methods by the role of the spin-spin correlations. To estimate the requirements to realize the room-temperature C-QED effects, in Sec. \ref{sec:CQED}, we study how the laser power controls the collective NV spins-microwave mode coupling, and the resulting Rabi oscillations and splitting. Our calculations indicate that due to population saturation effects the NV spin ensemble does not reach a sufficiently strong coupling with the microwave mode, and thus that a slightly larger number of NV spins is required to fully demonstrate the C-QED effects at room temperature. Finally, in Sec. \ref{sec:conclusions}, we summarize our conclusions and comment on possible extensions for future exploration.

\section{Multi-level Jaynes-Cumming Model \label{sec:multilevelJC}}


In this section, we present the multi-level JC model for the system shown in Fig. \ref{fig:sys}. We consider the quantum master equation for the reduced density operator $\hat{\rho}$ of the coupled NV centers-microwave resonator system:
\begin{align}
 & \partial_{t}\hat{\rho}=-\frac{i}{\hbar}\left[\hat{H}_{NV}+\hat{H}_{m}+\hat{H}_{m-NV} + \hat{H}_{m-m},\hat{\rho}\right]\nonumber \\
 & -\xi\sum_{k}\left(\mathcal{D}\left[\hat{\sigma}_{k}^{41}\right]\hat{\rho}+\mathcal{D}\left[\hat{\sigma}_{k}^{52}\right]\hat{\rho}+\mathcal{D}\left[\hat{\sigma}_{k}^{63}\right]\hat{\rho}\right)\nonumber \\
 & -\left(\xi+k_{sp}\right)\sum_{k}\left(\mathcal{D}\left[\hat{\sigma}_{k}^{14}\right]\hat{\rho}+\mathcal{D}\left[\hat{\sigma}_{k}^{25}\right]\hat{\rho}+\mathcal{D}\left[\hat{\sigma}_{k}^{36}\right]\hat{\rho}\right)\nonumber \\
 & -\sum_{k} (\sum_{i=4,5,6}k_{i7}\mathcal{D}\left[\hat{\sigma}_{k}^{7i}\right]\hat{\rho} + \sum_{i=1,2,3}k_{7i}\mathcal{D}\left[\hat{\sigma}_{k}^{i7}\right]\hat{\rho} )   \nonumber \\
 & -\sum_{k}\sum_{i=2,3}\left( k_{i1}\mathcal{D}\left[\hat{\sigma}_{k}^{1i}\right]\hat{\rho}+k_{1i}\mathcal{D}\left[\hat{\sigma}_{k}^{i1}\right]\hat{\rho} + 2 \chi_{i}\mathcal{D}\left[\hat{\sigma}_{k}^{ii}\right]\hat{\rho}\right) \nonumber \\
 & -\kappa\left[\left(n_{m}^{th}+1\right)\mathcal{D}\left[\hat{a}\right]\hat{\rho}+n_{m}^{th}\mathcal{D}\left[\hat{a}^{\dagger}\right]\hat{\rho}\right].\label{eq:me}
\end{align}
We consider the multiple levels of the $k$-th NV center as shown in Fig. \ref{fig:sys}(b), and denote the three spin levels (with projection $0,-1,+1$) of the triplet ground state $^{3}A_{2}$ as the levels $\left|1_{k}\right\rangle ,\left|2_{k}\right\rangle ,\left|3_{k}\right\rangle $, and those of the triplet excited state $^{3}E$ as the levels $\left|4_{k}\right\rangle ,\left|5_{k}\right\rangle ,\left|6_{k}\right\rangle $, and introduce the fictitious level $\left|7_{k}\right\rangle $ to represent the two singlet excited states. The NV centers are described by Hamiltonian $\hat{H}_{NV}=\hbar \omega_{31}\sum_{k=1}^{N}\hat{\sigma}_{k}^{31}\hat{\sigma}_{k}^{13} $ with the transition frequency $\omega_{31}$ between the $\left|1_{k}\right\rangle $ and $\left|3_{k}\right\rangle$ level. Here and in the following, we utilize the symbols $\hat{\sigma}_{k}^{ij}=\left|i_{k}\right\rangle \left\langle j_{k}\right|$ to present the projection operators (with $i=j$) and the transition operators (with $i\neq j$).  The transitions between other levels are not considered in the Hamiltonian, but through the dissipative super-operators as discussed later on. The microwave resonator is described by the Hamiltonian $\hat{H}_{m}=\hbar\omega_{m}\hat{a}^{\dagger}\hat{a}$ with the frequency $\omega_{m}$, the photon creation $\hat{a}^{\dagger}$ and annihilation operator $\hat{a}$, respectively. The energy exchange between the NV centers and the microwave resonator mode is described by the Hamiltonian $\hat{H}_{NV-m}=\hbar g_{31}\sum_{k}\left(\hat{a}^{\dagger}\hat{\sigma}_{k}^{13}+\hat{\sigma}_{k}^{31}\hat{a}\right)$ with the coupling strength $g_{31}$. The Hamiltonian $\hat{H}_{m-m}=\hbar\Omega \sqrt{\kappa/2} \hat{a} e^{i \omega_d t}+h.c.$ describes the driving of the microwave mode with a microwave field of frequency $\omega_d$ and driving strength $\Omega$, where $ \sqrt{\kappa/2}$ is the assumed transmission coefficient through the coupler (with the resonator decay rate $\kappa$). 

The remaining terms in Eq.(\ref{eq:me}) describe the system dissipation with Lindblad superoperator $\mathcal{D}\left[\hat{o}\right]\hat{\rho}=\frac{1}{2}\left\{ \hat{o}^{+}\hat{o},\hat{\rho}\right\} -\hat{o}\hat{\rho}\hat{o}^{+}$ for any operator $\hat{o}$. The second line describes the optical pumping with a rate $\xi$ from the spin levels of the triplet ground state to those of the triplet excited state. The third line describes the stimulated emission with the same rate $\xi$ and the spontaneous emission with a rate $k_{sp}$ from the spin levels of the triplet excited state to those of the triplet ground state.  For simplicity, we have ignored the much faster phonon relaxation process. The fourth line describes the inter-system cross with rates $k_{47},k_{67},k_{57}$ from the spin levels of the triplet excited state to the representative singlet excited state, and the similar process with rates $k_{73},k_{72},k_{71}$ from the singlet excited state to the spin levels of the triplet ground state. The fifth line describes the spin-lattice relaxation with rates $k_{31},k_{12}$ from the $+1,-1$ spin levels to the $0$ spin level on the triplet ground state, and with the rates $k_{13}\approx k_{31},k_{21}\approx k_{12}$ for the reverse processes, as well as the spin dephasing with the rates $\chi_{3},\chi_{2}$ of the $+1,-1$ spin levels. The spin-lattice relaxation is dominated by one photon process at extremely low temperature~\citep{TAstner}, and by two-phonon Orbach process and two-phonon Raman scattering around room temperature with $k_{31} \approx k_{12}$  ~\citep{Anorambuena,AJarmola}. To keep the model tractable, here, we utilize single dephasing rate to model qualitatively the spin dephasing due to the coupling with the spin environment, and the decoherence due to the inhomogeneous broadening of spin transitions. The last line describes the thermal photon emission and absorption of the resonator mode with the rate $\kappa$ and the thermal photon number $n_{m}^{th}=\left[e^{\hbar\omega_{m}/k_{B}T}-1\right]^{-1}$  at temperature $T$ (with the Boltzman constant $k_B$). The value of various rates and the calculation of $\xi$ from  the laser power $P$ are specified in the Appendix \ref{sec:params}.



To simulate trillions of NV centers, we solve the master equation (\ref{eq:me}) with the mean-field approach~\citep{DPlank}. In this approach, we derive the equation $\partial_{t}\left\langle \hat{o}\right\rangle =\mathrm{tr}\left\{ \hat{o}\partial_{t}\hat{\rho}\right\} $ for the expectation values $\left\langle \hat{o}\right\rangle =\mathrm{tr}\left\{ \hat{o}\hat{\rho}\right\} $ of any operator $\hat{o}$, and truncate the resulting equation hierarchy by approximating the mean values of products of many operators with those of few operators, i.e. the cumulant expansion approximation. In Eq. (\ref{eq:me}), we assume the same transition frequencies $\omega_{31}$, coupling strengths $g_{31}$ and rates $\xi,k_{sp},k_{ij},\chi_{2},\chi_{3}$ for all the NV centers, which allows us to utilize the symmetry raised to reduce dramatically the number of independent mean-field quantities. We utilize the QuantumCumulant.jl package~\citep{DPlank} to derive and solve the equations for the mean-field quantities up to second order, see the Appendix \ref{sec:codes}. In the Appendix \ref{sec:equations}, we consider the system without microwave driving, and present the derived equations for the population $\langle\hat{\sigma}_{1}^{ii}\rangle$ of the $i$-th level,  the mean intra-resonator photon number $\langle\hat{a}^\dagger \hat{a}\rangle$, the spin-photon correlation $\langle\hat{a}^\dagger \hat{\sigma}_{1}^{31} \rangle$ and the spin-spin correlation $\langle\hat{\sigma}_{1}^{31} \hat{\sigma}_{2}^{13}\rangle$. Here, the sub-index $1$ or $1,2$ indicate the mean-quantities for the first and second representative NV centers.


\section{Cooling of NV Spin Ensemble and Microwave Mode  \label{sec:modecooling}}

\begin{figure}[!htp]
\begin{centering}
\includegraphics[scale=0.50]{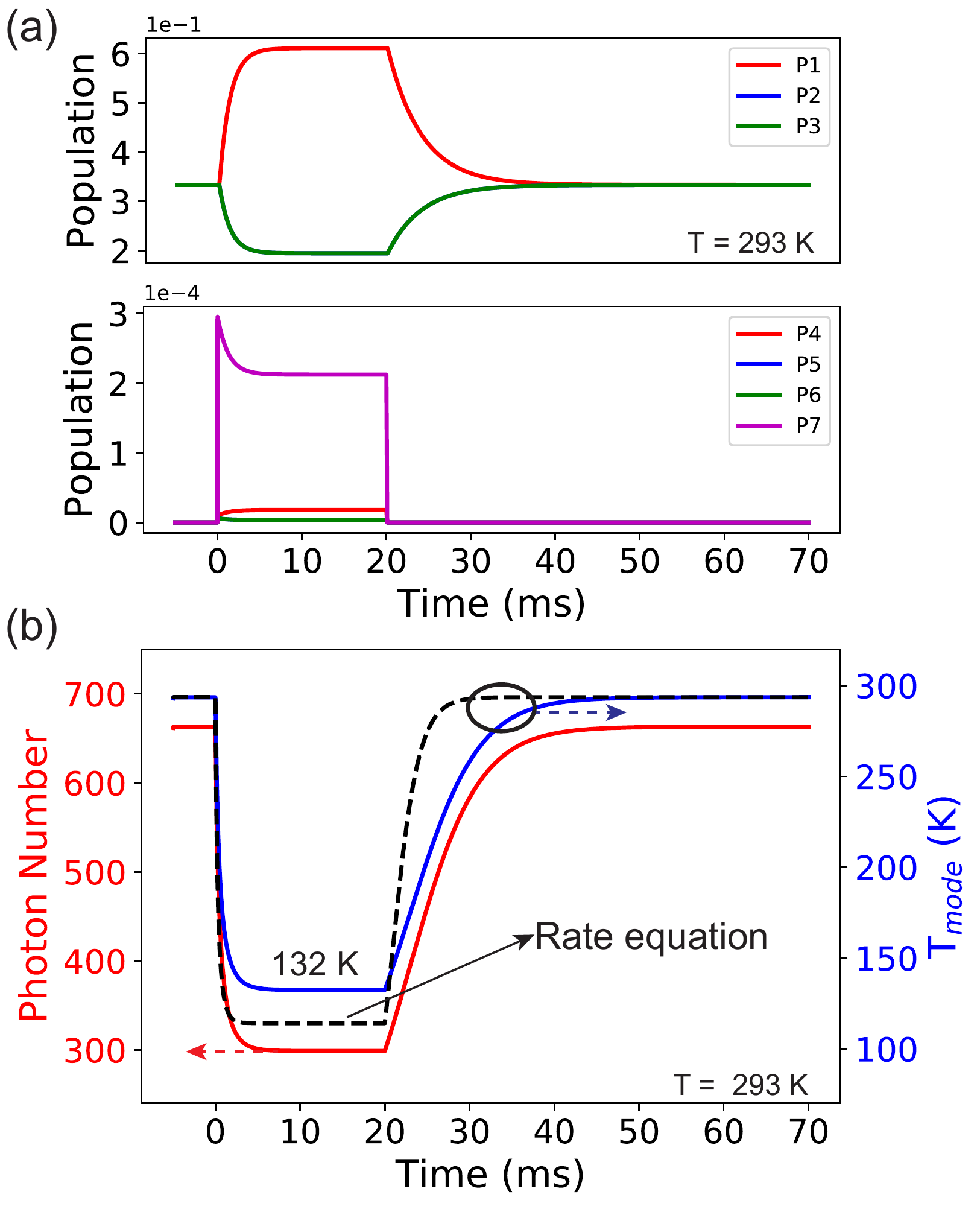}
\par\end{centering}
\caption{\label{fig:coolingdynamics-strong} Cooling dynamics of NV spin ensemble and resonator microwave mode. Panel (a) shows the population of the spin levels of the electronic ground state (the upper panel), and the population of those of the electronic excited state and the singlet excited state (lower panel). Panel (b) shows the intra-resonator photon number (red solid line, left axis) and the effective mode temperature (right axis) calculated with the JC model (blue solid line) and the rate equations (blue dashed line). Here, the diamond is illuminated by a laser with power $2$ W, and other parameters are specified in the Appendix \ref{sec:params}.}
\end{figure}

\begin{figure}[!htp]
\begin{centering}
\includegraphics[scale=0.5]{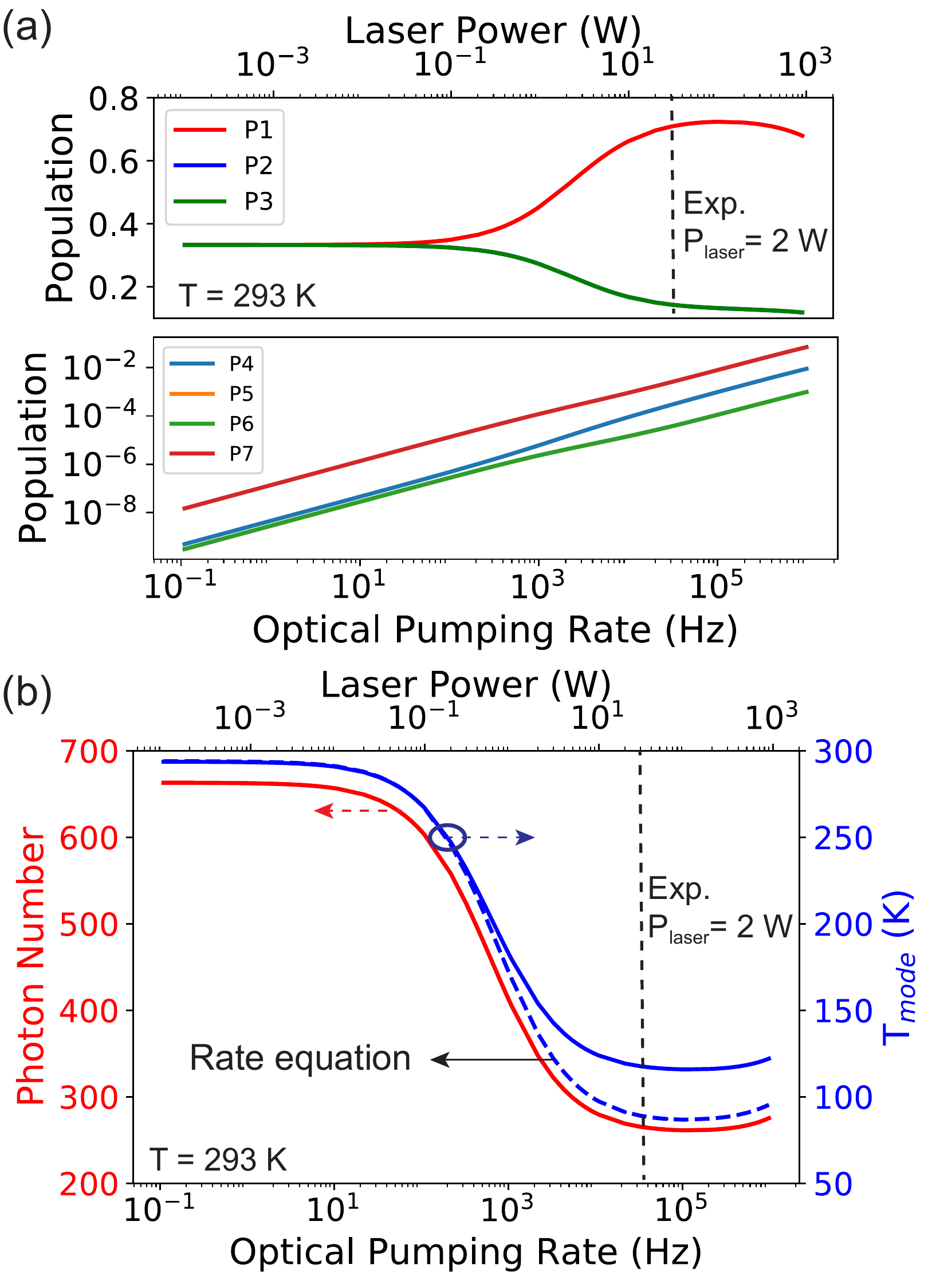}
\par\end{centering}
\caption{\label{fig:ste-state-cooling-strong} Steady-state cooling of the NV spins and the microwave resonator mode. Panel (a) shows the population of the spin levels of the electronic ground state (left axis of the upper panel), and the population of those of the electronic excited state and the singlet excited state (lower panel), as a function of  the optical pumping rate (lower axis) and the laser power (upper axis).  Panel (b) shows the intra-resonator photon number (red solid line, left axis) and the effective temperature (right axis) calculated with the multi-level JC model (blue solid line) and the rate equations (blue dashed line), as function of the optical pumping rate (lower axis) or the laser power (upper axis). The laser power $2$ W as used in Fig. \ref{fig:coolingdynamics-strong} and the experiment~\citep{NgW} is marked with the vertical dashed line. Other parameters are specified in the Appendix \ref{sec:params}.  }
\end{figure}

In this section, we demonstrate the optical spin cooling and the subsequent cooling of the microwave mode, by considering the system dynamics under a laser pulse excitation (Fig. \ref{fig:coolingdynamics-strong}), and the system steady-state under continuous laser excitation (Fig. \ref{fig:ste-state-cooling-strong}).  To quantify the non-equilibrium state of the spin ensemble and the resonator, we define the effective temperature  $T_{m}=\hbar\omega_{m}/\left[k_{B}\mathrm{ln}\left(1/\langle\hat{a}^{\dagger}\hat{a}\rangle+1\right)\right]$ for the microwave mode. A photon thermal bath at the effective temperature would lead to same intra-resonator photon number.

\subsection{Dynamics of Optical Spin Cooling and Microwave Mode Cooling}

Fig. \ref{fig:coolingdynamics-strong} shows the dynamics of the optical spin cooling and the subsequent microwave mode cooling. Before the laser excitation, the population of the three spin levels of the triplet ground state is about $\approx 1/3$ and that of the other levels is zero (Fig. \ref{fig:coolingdynamics-strong}a), and the intra-resonator photon number is around $661$ (Fig.\ref{fig:coolingdynamics-strong}b). The effective temperature of the microwave mode is equal to the room temperature $293$ K. 

When the laser excitation is on, the population of the $0$ spin level of the triplet ground state first increases and then reaches slowly to a constant value of $0.655$ (red curve in the upper part of Fig. \ref{fig:coolingdynamics-strong}a), and the populations of the $-1$ and $+1$ spin levels first decrease fast and then reach slowly to a constant value of $0.182$ and $0.162$, respectively (blue and green curve in the upper part of Fig. \ref{fig:coolingdynamics-strong}a). In addition, the population of the spin levels of the triplet excited state increases dramatically and then behaves similarly as that of the spin levels of the triplet ground state except for the five orders of magnitude smaller value (curves in the lower part of Fig. \ref{fig:coolingdynamics-strong}a), and the population of the singlet excited state behaves similar as the upper spin levels of the triplet excited state except for the fifty times larger value (upmost curve in the lower part of Fig. \ref{fig:coolingdynamics-strong}a). In the meanwhile, the photon number drops dramatically and then decays slowly to a constant value below $298$ within 20 ms, which leads to a reduction of the effective microwave mode temperature from $293$ K to $132$ K, about a reduction of $161$ K (Fig. \ref{fig:coolingdynamics-strong}b), which is about $56$ K smaller than the value reported previously~\citep{NgW}.  

When the laser excitation is off, the populations of the $\pm1$ and $0$ spin level of the triplet ground state increase and decrease slowly to the initial values within about $20$ ms (upper part of Fig. \ref{fig:coolingdynamics-strong}a). In addition, the population of other levels drops dramatically to zero (lower part of Fig. \ref{fig:coolingdynamics-strong}a). In the meanwhile, the photon number increases slowly to the initial value, which leads to the increase of the effective mode temperature to the room temperature (Fig. \ref{fig:coolingdynamics-strong}b). These results are similar to those in the experiment~\citep{NgW} (see the Appendix \ref{sec:optcoolinglowfrequency}).

To gain more insights into the optical spin cooling and the subsequent microwave mode cooling, we analyze further the mean-field equations. To verify whether the spin-spin correlation $\langle\hat{\sigma}_{1}^{31} \hat{\sigma}_{2}^{13}\rangle$ plays a role in the cooling, we neglect this correlation $\langle\hat{\sigma}_{1}^{31} \hat{\sigma}_{2}^{13}\rangle \approx 0 $,  and then eliminate adiabatically the spin-photon correlation $\langle\hat{a}^\dagger \hat{\sigma}_{1}^{31} \rangle$ to arrive at rate equations for the population of the NV levels and the intra-resonator photon number, see the Appendix \ref{sec:rateeqn}. These rate equations can reproduce qualitatively the dynamics shown in Fig. \ref{fig:coolingdynamics-strong} except that the photon number reduces to $257$,  which corresponds to the effective mode temperature of $114$ K (see the blue dashed line in Fig. \ref{fig:coolingdynamics-strong}b), which is about $18$ K smaller than the actual value. This indicates that the spin-spin correlation does play a role here. Note that for the setup in the deep weak coupling regime as in the experiment~\citep{NgW}, the rate equations can reproduce the results from the multi-level Jaynes-Cumming model (see the Appendix \ref{sec:optcoolinglowfrequency}).

\subsection{Steady-state Optical Spin Cooling and Microwave Mode Cooling}

After understanding the cooling dynamics, we study now the cooling performance at steady-state for different optical pumping rate (laser power), see Fig.\ref{fig:ste-state-cooling-strong}. We find that the populations of the $0$ and $\pm1$ spin levels of the triplet ground state increase and decrease when the optical pumping rate (laser power) exceeds about $10$ Hz ($9.1\times 10^{-3}$ W), and become saturated at the values about $0.73$ and $0.13$ for the optical pumping rate about $10^5$ Hz (about $100$ W), but start decreasing for much larger optical pumping rates (laser power), respectively, see the upper part of Fig.\ref{fig:ste-state-cooling-strong}a. Furthermore, we find that the populations of the triplet and singlet excited levels increase always with increasing optical pumping rate, and the latter is always larger than the former, and the population of the $0$ spin level of the triplet excited state becomes larger than that of the $\pm1$ spin levels once the optical pumping rate (laser power) exceeds $10$ Hz ($9.1\times 10^{-3}$ W), see the lower part of Fig. \ref{fig:ste-state-cooling-strong}a.

Accompanying the change of population, the intra-resonator photon number decreases dramatically for the optical pumping rate (laser power) exceeding $10$ Hz ($9.1\times10^{-3}$ W), and approaches the minimum about $261$ for the optical pumping rate (laser power) about $10^5$ Hz ($10^2$ W), and finally start increasing for much larger optical pumping rate, see Fig.\ref{fig:ste-state-cooling-strong}b. Converting the photon number to the effective temperature, we find that the mode temperature decreases from the room temperature ($293$ K) to the minimum around $116$ K (a reduction of $177$ K), which is $72$ K lower than the value achieved so far~\citep{NgW}. Same as before, the effective mode temperature calculated with the rate equations is about $29$ K lower than the actual value (blue dashed line in Fig.\ref{fig:ste-state-cooling-strong}b), indicating again the influence of the spin-spin correlation on the microwave mode cooling in the current system. If we manage to increase the number of NV centers to the same level  $1.6 \times 10^{15}$ as in the experiment~\citep{NgW} (about $40$ times enhancement), the intra-resonator photon number reduces to $71$, corresponding to an effective temperature of $32$ K (not shown). This simple calculation highlights the perspective of further improving the microwave mode cooling with more NV spins.

\begin{figure}[!htp]
\begin{centering}
\includegraphics[scale=0.5]{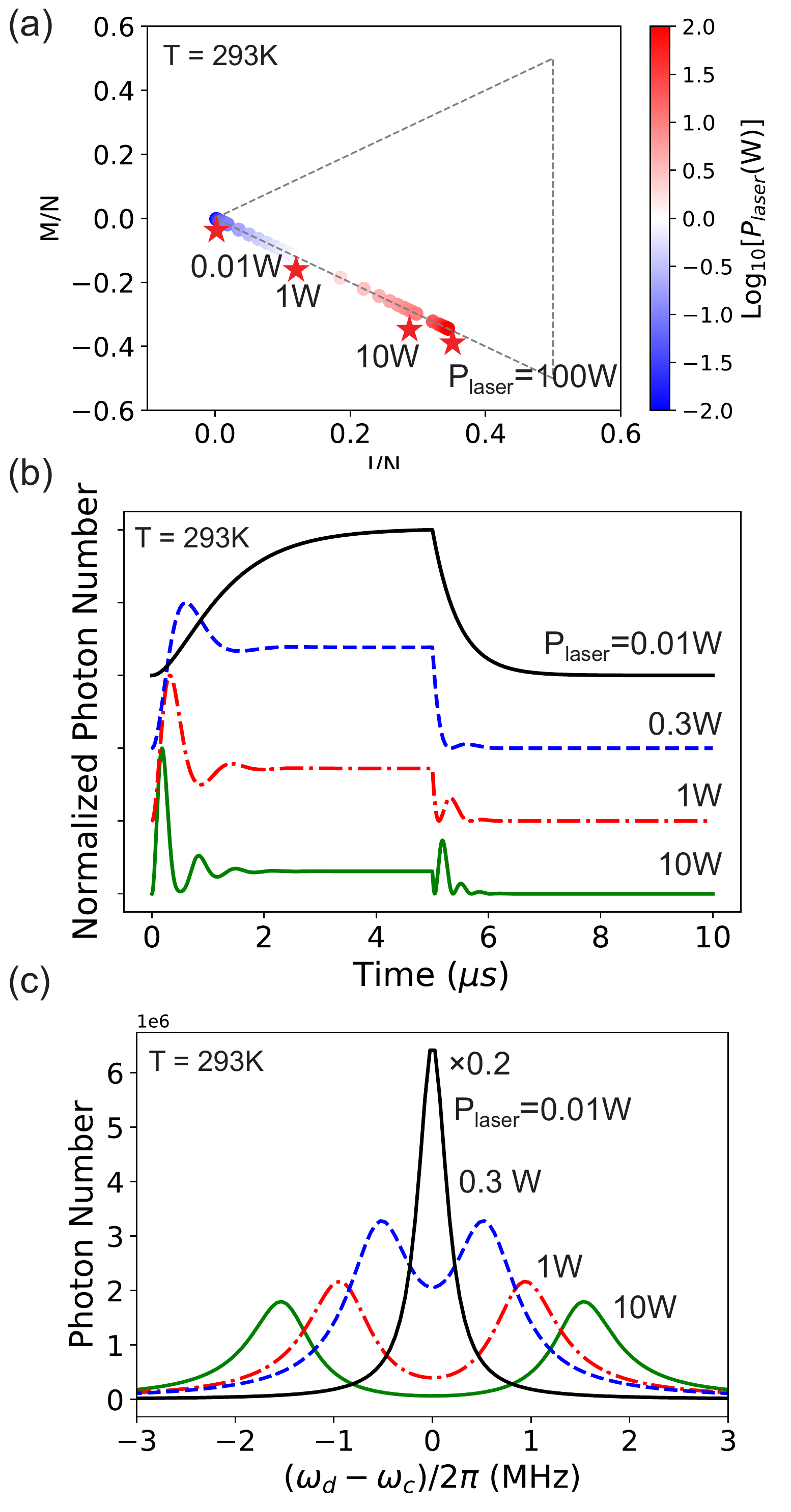}
\par\end{centering}
\caption{\label{fig:CQED} Room-temperature C-QED effects with multi-level NV centers coupled to the $532$ nm laser field and the microwave resonator. Panel (a) transforms the $+1,0$ spin-level population into the average of the Dicke state quantum numbers $M,J$ (normalized by the total number $N$ of the NV centers) for the increasing laser power. The gray dashed lines indicate the boundary of the Dicke state space. Panel (b) shows the dynamics of the normalized intra-resonator photon number for the laser power $0.01,0.3,1,10$ W (black solid, blue dashed, red dash-dotted and green solid line), where the curves are vertically shifted for the sake of clarity. Panel (c) shows the steady-state intra-resonator photon number as function of the detuning of the microwave driving field frequency $\omega_d$ and the microwave resonator frequency $\omega_c$ for the laser power $0.01,0.3,1,10$ W(black solid, blue dashed, red dash-dotted and green solid line). The microwave field driving amplitude is $\Omega = 2\pi \times 9.7\times10^{5}$, $2\pi\times9.7\times10^{8}$ Hz$^{-1/2}$ for the panel (b) and (c), respectively.  }
\end{figure}

\section{Room Temperature C-QED Effects \label{sec:CQED}}

In the previous section, we demonstrated that the optically cooled NV spin ensemble can be used to cool the microwave resonator mode. In this section, we demonstrate that it can be also utilized to realize the collective strong coupling with the cooled microwave mode, and to manifest the cavity-QED effects at room temperature. These effects have been studied previously by us~\citep{zhangyuan1} with a model, which treats the NV spins as two-level systems and the optical spin cooling effectively with single spin relaxation rate. Here, with the more advanced model, we are able to estimate the laser power required to observe these C-QED effects, which can guide directly the experimental research in future.

To proceed, we firstly translate the spin population polarization into the average of the Dicke states numbers $J,M$ to illustrate more intuitively the collective coupling strength. In our previous study ~\citep{zhangyuan1}, we show that these averaged numbers can be calculated as $M=J_0(2p-1),J(J+1)=(2p-1)^2J_0(J_0+1)+6p(p-1)J_0$ with $J_0=N/2$~\citep{WesenbergJ} with the population $p$ on the upper spin level coupled resonantly to the microwave resonator. For the current system with multiple levels, we calculate the relative population $p=\langle\hat{\sigma}_{1}^{33}\rangle/(\langle\hat{\sigma}_{1}^{11}\rangle +\langle\hat{\sigma}_{1}^{33}\rangle)$ with  the population $\langle\hat{\sigma}_{1}^{11}\rangle,\langle\hat{\sigma}_{1}^{33}\rangle$ calculated with the multi-level JC model, and then the averaged numbers $J,M$. In Fig. \ref{fig:CQED} (a), we plot the evolution of these numbers normalized to the number of NV centers $N$ with increasing laser power in the Dicke states space enclosed by the dashed lines.  We see that as the laser power increases from $10^{-2}$ W to $1$ W and finally to $100$ W, these numbers start from the leftmost corner for the Dicke states with smaller $J$ or lower symmetry, and evolve along the lower boundary for the Dicke states with larger $J$ or higher symmetry, and finally end up at the Dicke states with $J/N\approx0.35$ due to the saturation effect as described above, which is still far away from the rightmost corner for the Dicke state with largest $J$ and highest symmetry.

Since the NV spin-microwave mode coupling scales as $\propto \sqrt{2J} g$, we expect that the coupling increases with laser power. For the current system, we estimate the largest coupling strength as $\sqrt{2J}g\approx 2\pi\times 0.57$ MHz, and conclude that the system is on the edge of the strong coupling regime since this strength is slightly smaller than the spin-dephasing rate $\chi_3=2\pi\times 0.64$ MHz. Thus, to demonstrate the C-QED effects at room temperature, we have to increase the coupling strength by either increasing the number of NV centers or optimizing the single spin-microwave mode coupling. As an example, in the following, we consider the system with  ten times more NV spins, i.e. $N=4\times 10^{14}$, which might be achieved with better diamond sample, to obtain the collective coupling $\sqrt{2J}g\approx2\pi\times 1.8$ MHz, which is now larger than the spin-dephasing rate and thus brings the system into the strong coupling regime.  

For the system with more spins, we study firstly the Rabi oscillations, see Fig.\ref{fig:CQED}(b). Here, we apply initially the laser illumination with increasing power $0.01,0.1,1,10$ W to cool the NV spin ensemble and the microwave mode, and then drive the microwave resonator resonantly with a microwave field of given amplitude for $5$ ${\mu s}$, and finally study the dynamics of the intra-resonator photon number. For the smallest laser power $0.01$ W, the photon number increases firstly with time and then saturates when the driving field is on, and decreases exponentially to some finite value when the driving field is off. For larger laser power $0.1$ W, the photon number shows a bump before reaching the saturation and the finite value. When the laser power increases further to $1$ W and $10$ W, the bump evolves to oscillations, and the oscillation becomes slightly faster. These results are caused by the transition from the weak to strong collective coupling due to the increased coupling strength. 

Then, we investigate the Rabi splittings, see Fig.\ref{fig:CQED}(c). Here, we follow the similar procedure as before except that we drive the microwave resonator continuously with a microwave field of fixed amplitude, and then study the intra-resonator photon number at steady-state as function of the frequency detuning of the driving field and the microwave resonator. We find that the photon number shows a single peak at zero frequency detuning for the smallest laser power $0.03$ W, and this single peak evolves into two split peaks with weaker strength for larger laser power $0.1$ W, and these split peaks become further apart and weaker for much larger laser power $1$ and $10$ W. For even larger laser power $100$ W, the photon number is similar as that for the laser power $10$ W due to the cooling saturation as described above. 


\section{Conclusions \label{sec:conclusions}}

In summary, we have proposed to utilize a setup similar to that in the diamond maser experiment~\citep{Breeze}, which features a high frequency microwave resonator of lower damping rate coupled strongly to the NV spin ensemble of lower spin-dephasing rate, to realize better  microwave mode cooling and room temperature C-QED effects (Rabi oscillation and splitting) with the optically cooled diamond NV spin ensemble. We have developed a multi-level Jaynes-Cumming (JC) model to describe precisely the optical spin cooling and the collective NV spins-resonator coupling. By modifying this model, one can also describe systems with other solid-state spin systems and investigate other interesting effects, such as pulsed and continuous-wave masing. 

Our calculations predict the microwave mode cooling down to $116$ K, about $72$ K lower than the record achieved so far with the diamond NV ensembles~\citep{NgW}, and this value is also about $29$ K larger than the value calculated with rate equations, highlighting the importance the spin-spin correlation. Our calculations show also that the collective coupling with the microwave mode  can be dramatically enhanced for the optically cooled NV spin ensemble, as clearly manifested with the increased average of the Dicke state quantum numbers for large laser power. For the system with slightly more NV spins, the system can reach the strong coupling regime for the moderate laser power, which leads to the realization of the Rabi splittings and oscillations at room temperature. In addition, we estimate the influence of the optical heating of the diamond on the microwave mode cooling and the C-QED effects in the Appendix \ref{sec:opticalheating}, but conclude that this influence can be mitigated by cooling the diamond with various techniques. We note that, accompanying with the reprint of this work, 
an experiment~\citep{DAFahey} reported the observation of laser-power controlled Rabi splittings with NV centers at room temperature. 


\section*{Author contributions}
Yuan Zhang and Qilong Wu contribute equally to this work. All the authors contribute to the writing of the manuscript.

\begin{acknowledgments}
This work was supported by the National Natural Science Foundation of China project No. 12004344, 62027816, and Henan Center for Outstanding Overseas Scientists project GZS201903, as well as the Danish National Research Foundation through the Center of Excellence for Complex Quantum Systems (Grant agreement No. DNRF156), and the European Union's Horizon 2020 Research and Innovation Programme under the Marie Sklodowska-Curie program (No. 754513).
\end{acknowledgments}

\appendix

\begin{table}
\begin{centering}
\begin{tabular}{|c|c|c|}
\hline 
Description & Symbol & Value\tabularnewline
\hline
\hline
\multicolumn{3}{|c|}{ Transition rates between NV multiple levels~\citep{NgW}} \tabularnewline
\hline
Optical pumping and decay rate & $\xi$ & $0-1$ MHz\tabularnewline
\hline 
Spontaneous emission rate & $k_{sp}$ & $66$ MHz\tabularnewline
\hline 
Inter-system crossing rate & $k_{47}$ & $7.9$ MHz\tabularnewline
\hline 
 & $k_{67}=k_{57}$ & $53$ MHz\tabularnewline
\hline 
 & $k_{73}=k_{72}$ & $0.73$ MHz\tabularnewline
\hline 
 & $k_{71}$ & $1.0$ MHz\tabularnewline
 \hline 
\multicolumn{3}{|c|}{Parameters related to the diamond maser experiment~\citep{Breeze}} \tabularnewline
\hline 
Resonator Frequency & $\omega_{c}$ & $2\pi\times9.22$ GHz\tabularnewline
\hline 
Photon damping rate & $\kappa$ & $1.88$ MHz\tabularnewline
\hline 
Spin-resonator coupling & $g_{31}$ & $0.69$ Hz\tabularnewline
\hline 
Number of spins & $N$ & $4\times10^{13}$\tabularnewline
\hline 
Spin transition frequency & $\omega_{31}$ & $2\pi\times 9.22$ GHz \tabularnewline
\hline 
  & $\omega_{12}$ & $2\pi\times 3.48$ GHz \tabularnewline
\hline 
Spin dephasing rate  & $\chi_{2}=\chi_{3}$ & $2\pi\times 0.64$ MHz \tabularnewline
\hline
Spin-lattice relaxation rate & $k_{31}\approx k_{12}$ & $208$ Hz\tabularnewline
\hline
\end{tabular}
\par\end{centering}
\caption{\label{tab:param}Parameters for the dissipation rates for NV centers (upper part), as reported ~\citep{NgW}, and the parameters related the microwave resonator, as reported in the diamond maser experiment~\citep{Breeze}. In our proposed setup, the $+1 \to 0$ spin transition with frequency $\omega_{31}$ couples resonantly to the microwave resonator mode.  }
\end{table}

\section{\label{sec:params} Parameters for the Simulations in the Main Text}

In Tab. \ref{tab:param}, we present the parameters utilized in the simulations of the main text, where the upper part is from the microwave mode cooling experiment~\citep{NgW} and the lower part is mainly from the diamond maser experiment~\citep{Breeze}. Here, we assume the relatively smaller Zeeman shift induced by a relatively weak magnetic field so that the $+1 \to 0$ spin transition couple resonantly with the microwave resonator.  

According to the experiment~\citep{NgW}, the optical pumping rate $\xi$ can be calculated with the following expression
\begin{equation}
\xi=\frac{\lambda_{P}\sigma_{\lambda_{P}}}{hcA_{p}l\alpha}\left(1-e^{-l\alpha}\right)\left(1-R\right)P.
\end{equation}
Here, $\sigma_{\lambda_{p}}=3.1\times10^{-21}\mathrm{m}^{2}$ is the absorption cross-section of the NV centers at the laser wavelength $\lambda_{p}=532$ nm. $h=6.63\times10^{-34}$ $\mathrm{J \cdot s}$ is the Planck constant, $c=2.99\times10^{8}$ $\mathrm{m/s}$ is the speed of light. $A_{p}=1.76\times10^{-6}$ $\mathrm{m^{2}}$ is the cross-sectional area of the pump beam on the sample. $l=1.5$ mm is the thickness of the diamond crystal. $\alpha=2.3\times10^{3}$ $\mathrm{m}^{-1}$ is the absorption coefficient at the pumping laser wavelength $\lambda_{P}$. $R=\left|\frac{n_{1}-n_{2}}{n_{1}+n_{2}}\right|^2$ is Fresnel reflection
coefficient with the refractive index of air $n_{1}=1$ and of the diamond $n_2 =2.42$. $P$ is the laser power in unit of W. These parameters are estimated for the setup in Ref. ~\citep{NgW}, and might vary for the proposed setup. In any case, we expect that these values might lead to a reasonable estimation of the optical pumping rate for given laser power.

It is worth to noting that by optimizing the system setup, such as the absorption of the diamond with taped configuration~\citep{WuH} or the total reflection~\citep{ClevensonH}, the optical pumping rate $\xi$ might be increased by one to one to two orders of magnitude for the same laser power. Furthermore, the typical commercial CW solid-state laser of $532$ nm can provide the power up to $20$ W, and the pulsed solid-state laser can provide the peak power up to $100$ W or even $10^3$ W.

\begin{figure}[!htp]
\begin{centering}
\includegraphics[scale=0.46]{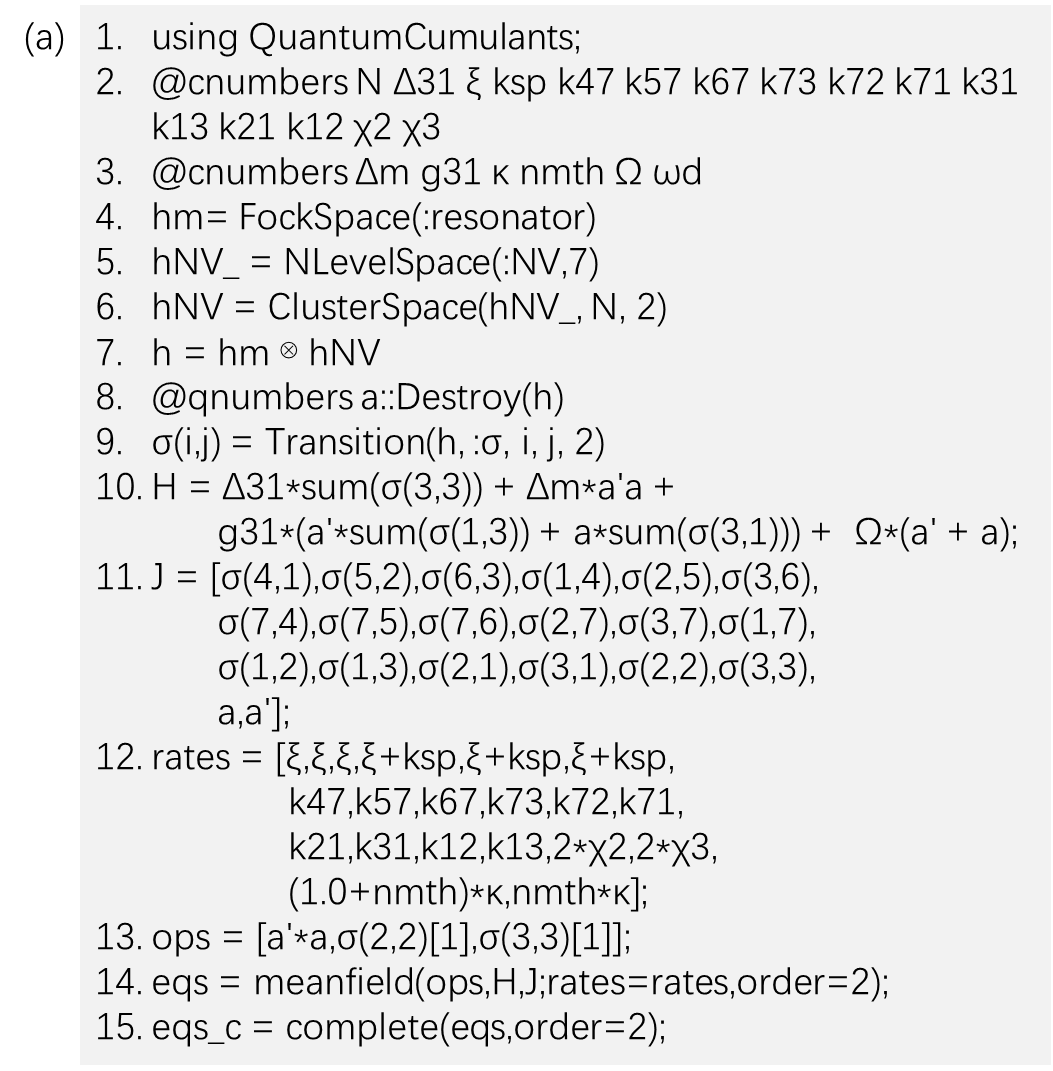}
\includegraphics[scale=0.42]{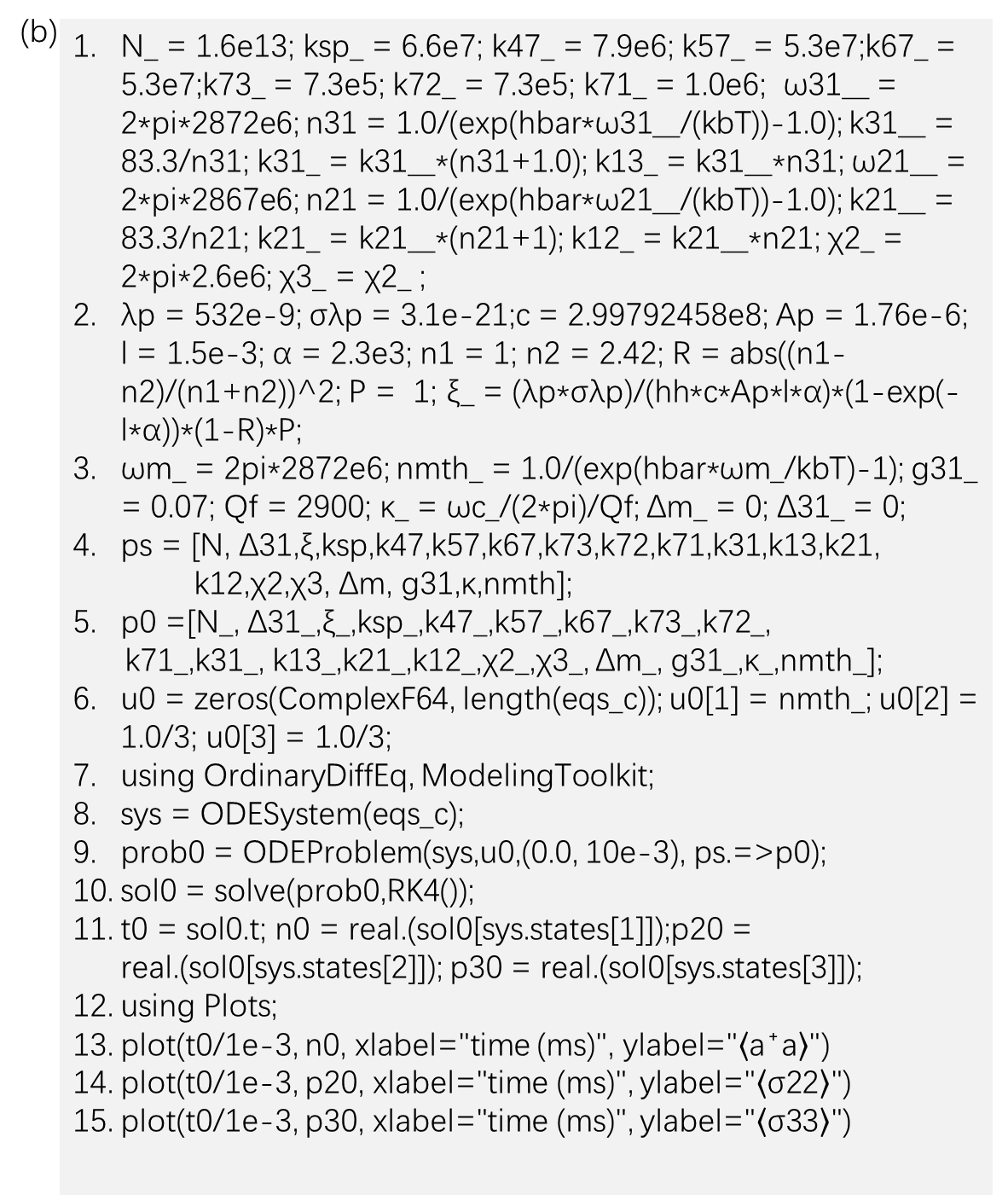} 
\par\end{centering}
\caption{\label{fig:codes}Julia codes to derive the mean-field equations (a) and to solve the equations (b).}
\end{figure}

\section{Julia Codes to Derive and Solve Mean-field Equations\label{sec:codes}}
In this Appendix, we present the Julia code to derive the mean-field equations (Fig.\ref{fig:codes}a) and to solve these equations (Fig.\ref{fig:codes}b). First, we explain the former code. The 1st line imports the QuantumCumulants.jl package, and the 2nd and 3rd lines define the complex numbers. Here, we work in the  frame rotating with the microwave field frequency $\omega_d$ and introduce the frequency detuning $\Delta \omega_{31}=\omega_{31}-\omega_d$ and $\Delta \omega_m = \omega_m -\omega_d$. The 4th line defines the Hilbert space of the microwave resonator as a quantized harmonic oscillator, while the 5th line defines that of the single NV center as a seven-levels system. The 6th line defines the Hilbert space of the $N$ NV centers, where the last argument indicates the number of representative operators labeled by $k=1,2$, and the 7th line defines the Hilbert space of the NV centers-microwave resonator system. The 8th line defines the annihilation operator of the microwave photons, and the 9th line defines the transition operators $\sigma\left(i,j\right)=\left\{ \hat{\sigma}_{1}^{ij},\hat{\sigma}_{2}^{ij}\right\} $
with $i\neq j$ and the projection operators $\sigma\left(i,i\right)=\left\{ \hat{\sigma}_{1}^{ii},\hat{\sigma}_{2}^{ii}\right\} $. The 10th line defines the system Hamiltonian, while the 11th and 12th line define the list of operators and values to specify the Lindblad super-operators. The 13th line defines a list of three operators, and the 14th line derives the equations for the expectation value of these operators. The 15th line analyzes the unknown mean-field quantities, and derives the equations for them to form a closed set of equations, see the Appendix \ref{sec:equations}. In the derivation, we have assumed vanishing third-order cumulant to approximate $\left\langle  \hat{o}\hat{p}\hat{q}\right\rangle $
as $\left\langle \hat{o}\right\rangle \left\langle \hat{p}\hat{q}\right\rangle +\left\langle \hat{p}\right\rangle \left\langle \hat{o}\hat{q}\right\rangle +\left\langle \hat{q}\right\rangle \left\langle \hat{o}\hat{p}\right\rangle -2\left\langle \hat{o}\right\rangle \left\langle \hat{p}\right\rangle \left\langle \hat{q}\right\rangle $
for any operators $\hat{o},\hat{p},\hat{q}$. 

Second, we describe the code to solve the equations, see Fig.\ref{fig:codes} (b). The 1st line specifies the parameters related to the NV centers, and the 2nd line calculates the optical pumping rate for the laser excitation with the reference power $1$ W. The 3rd line specifies the parameters related to the microwave resonator.  The 4th and 5th line define the list of the symbols and of their values. The 6th line defines the initial conditions for the unknown mean-field quantities. The 7th line imports the  "OrdinaryDiffEq" and "ModelingToolkit" packages to define and solve the ordinary differential equations (ODE). The 8th line defines the ODE system with the derived mean-field equations, the 9th line defines the ODE problem with the initial condition, the time argument and the parameters. The 10th line solves the ODE with Rounge-Kutta method, and the 11th line extracts the list of simulation time and the physical quantities of interest. The 12th line imports the "Plots" package, and the remaining lines  plot the dynamics of the photon number and the population. 

\section{Closed Set of Second-Order Mean-Field Equations for Microwave Mode Cooling \label{sec:equations}}

In this Appendix, we show the derived closed set of mean-field equations in second order. Here, we focus on the microwave mode cooling, and do not include the microwave field driving. First, we present the equations for the first-order mean-field quantities. The population $\langle\hat{\sigma}_{1}^{ii}\rangle$ of various levels of NV centers follows the equations
\begin{align}
 & \partial_{t}\langle\hat{\sigma}_{1}^{11}\rangle =  - (k_{12}+k_{13} + \xi )\langle\hat{\sigma}_{1}^{11}\rangle + (\xi+k_{sp}) \langle\hat{\sigma}_{1}^{44}\rangle \nonumber \\
 & + k_{21}\langle\hat{\sigma}_{1}^{22}\rangle+k_{31}\langle\hat{\sigma}_{1}^{33}\rangle -ig_{31}\left(\langle\hat{a}^{\dagger}\hat{\sigma}_{1}^{13}\rangle-\langle\hat{a}\hat{\sigma}_{1}^{31}\rangle\right), \label{eq:11} \\
 & \partial_{t}\langle\hat{\sigma}_{1}^{22}\rangle=-\xi\langle\hat{\sigma}_{1}^{22}\rangle+(k_{sp}+\xi)\langle\hat{\sigma}_{1}^{55}\rangle\nonumber \\
 & +k_{72}\langle\hat{\sigma}_{1}^{77}\rangle+k_{12}\langle\hat{\sigma}_{1}^{11}\rangle-k_{21}\langle\hat{\sigma}_{1}^{22}\rangle, \label{eq:22} \\
 & \partial_{t}\langle\hat{\sigma}_{1}^{33}\rangle=-\xi\langle\hat{\sigma}_{1}^{33}\rangle+(k_{sp}+\xi)\langle\hat{\sigma}_{1}^{66}\rangle\nonumber \\
 & +k_{73}\langle\hat{\sigma}_{1}^{77}\rangle+k_{13}\langle\hat{\sigma}_{1}^{11}\rangle-k_{31}\langle\hat{\sigma}_{1}^{33}\rangle\nonumber \\
 & +ig_{31}\left(\langle\hat{a}^{\dagger}\hat{\sigma}_{1}^{13}\rangle-\langle\hat{a}\hat{\sigma}_{1}^{31}\rangle\right), \label{eq:33} \\
&\partial_{t} \langle\hat{\sigma}_{1}^{44}\rangle = \xi\langle\hat{\sigma}_{1}^{11}\rangle  -(k_{sp}+\xi)\langle\hat{\sigma}_{1}^{44}\rangle-k_{47}\langle\hat{\sigma}_{1}^{44}\rangle, \label{eq:44} \\
&\partial_{t}\langle\hat{\sigma}_{1}^{55}\rangle=\xi\langle\hat{\sigma}_{1}^{22}\rangle-(\xi+k_{sp})\langle\hat{\sigma}_{1}^{55}\rangle-k_{57}\langle\hat{\sigma}_{1}^{55}\rangle, \label{eq:55} \\
&\partial_{t}\langle\hat{\sigma}_{1}^{66}\rangle=\xi\langle\hat{\sigma}_{1}^{33}\rangle-(\xi+k_{sp})\langle\hat{\sigma}_{1}^{66}\rangle-k_{67}\langle\hat{\sigma}_{1}^{66}\rangle, \label{eq:66} \\
 & \partial_{t}\langle\hat{\sigma}_{1}^{77}\rangle=k_{47}\langle\hat{\sigma}_{1}^{44}\rangle+k_{57}\langle\hat{\sigma}_{1}^{55}\rangle\nonumber \\
 & +k_{67}\langle\hat{\sigma}_{1}^{66}\rangle-\left(k_{71}+k_{72}+k_{73}\right)\langle\hat{\sigma}_{1}^{77}\rangle. \label{eq:77}
\end{align}

Second, we illustrate the equations for the second-order mean values. The mean intra-resonator photon number $\langle\hat{a}^{\dagger}\hat{a}\rangle$ satisfies
the equation
\begin{equation}
 \partial_{t}\langle\hat{a}^{\dagger}\hat{a}\rangle=\kappa\left(n_{m}^{th}-\langle\hat{a}^{\dagger}\hat{a}\rangle\right) +iNg_{31}\left(\langle\hat{a}\hat{\sigma}_{1}^{31}\rangle-\langle\hat{a}^{\dagger}\hat{\sigma}_{1}^{13}\rangle\right). \label{eq:photon}
\end{equation}
The NV-photon correlation $\langle\hat{a}^{\dagger}\hat{\sigma}_{1}^{13}\rangle$
follows the equation
\begin{align}
 & \partial_{t}\langle\hat{a}^{\dagger}\hat{\sigma}_{1}^{13}\rangle=i\left(\omega_{m}-\omega_{31}\right)\langle\hat{a}^{\dagger}\hat{\sigma}_{1}^{13}\rangle\nonumber \\
 & -\left[\left(\kappa+k_{12}+k_{13}+k_{31}\right)/2+\xi+\chi_{3}\right]\langle\hat{a}^{\dagger}\hat{\sigma}_{1}^{13}\rangle\nonumber \\
 & +ig_{31}\left[\langle\hat{\sigma}_{1}^{33}\rangle\left(1+\langle\hat{a}^{\dagger}\hat{a}\rangle\right)-\langle\hat{\sigma}_{1}^{11}\rangle\langle\hat{a}^{\dagger}\hat{a}\rangle\right]\nonumber \\
 & +i(N-1)g_{31}\langle\hat{\sigma}_{1}^{31}\hat{\sigma}_{2}^{13}\rangle. \label{eq:correlation}
\end{align}
The correlations $\langle\hat{a}\hat{\sigma}_{1}^{31}\rangle$
are simply the complex conjugation of $\langle\hat{a}^{\dagger}\hat{\sigma}_{1}^{13}\rangle$.
In the end, the NV-NV correlation $\langle\hat{\sigma}_{1}^{31}\hat{\sigma}_{2}^{13}\rangle$
satisfies the equation 
\begin{align}
 & \partial_{t}\langle\hat{\sigma}_{1}^{31}\hat{\sigma}_{2}^{13}\rangle=-\left[k_{12}+k_{13}+k_{31}+2\left(\xi+\chi_{3}\right)\right]\langle\hat{\sigma}_{1}^{31}\hat{\sigma}_{2}^{13}\rangle\nonumber \\
 & +ig_{31}\left(\langle\hat{\sigma}_{1}^{33}\rangle-\langle\hat{\sigma}_{1}^{11}\rangle\right)\left(\langle\hat{a}\hat{\sigma}_{1}^{31}\rangle-\langle\hat{a}^{\dagger}\hat{\sigma}_{1}^{13}\rangle\right).
\end{align}

\begin{table}
\begin{centering}
\begin{tabular}{|c|c|c|}
\hline 
Frequency & $\omega_{c}=2\pi f_{c}$ & $2\pi\times2.872$ GHz\tabularnewline
\hline 
Quality-factor & $Q=f_{c}/\kappa$ & $2900$\tabularnewline
\hline 
Damping rate & $\kappa$ & $6.22$ MHz\tabularnewline
\hline 
Spin-resonator coupling & $g_{31}$ & $0.084$ Hz\tabularnewline
\hline 
Number of spins & $N$ & $1.6\times10^{15}$\tabularnewline
\hline 
Spin transition frequency & $\omega_{31}$ & $2\pi\times 2.872$ GHz \tabularnewline
\hline 
  & $\omega_{21}$ & $2\pi\times 2.867$ GHz \tabularnewline
\hline
Spin-lattice relaxation rate & $k_{31}\approx k_{12}$ & $ 83$ Hz\tabularnewline
\hline 
Spin dephasing rate  & $\chi_{2}=\chi_{3}$ & $2\pi\times2.6$ MHz \tabularnewline
\hline
\end{tabular}
\par\end{centering}
\caption{\label{tab:param-cooling} Parameters related to the microwave mode cooling~\citep{NgW}. }
\end{table}

\section{\label{sec:optcoolinglowfrequency} Microwave Mode Cooling with a Lower Frequency Resonator}

In this Appendix, we present the results similar as in the main text but for the setup used in the microwave mode cooling experiment~\citep{NgW}, which features a microwave resonator with lower frequency. In Tab. \ref{tab:param-cooling}, we present the parameters utilized for the simulations. Here, we assume the spin dephasing rate $2\pi \times 2.6$ MHz, which is consistent with the value reported in the experiments on the C-QED effects at low temperature~\citep{SPutz,KuboY}, but orders of magnitude larger than $0.33$ MHz as estimated in ~\citep{NgW}. We note the underestimation of this parameter in~\citep{NgW} is because they assumed that the ${\rm NV}^{-}$ concentration is equal to that of the $N_s^0$ defects (also known as P1 centers), but usually the former is orders of magnitude smaller than the latter~\citep{TLuo}. In addition, we note that the value we used is in the same order of magnitude as the width of the zero-field transient electron paramagnetic resonance measurement, see Fig. 2(b) in~\citep{NgW}.

\begin{figure}
\begin{centering}
\includegraphics[scale=0.5]{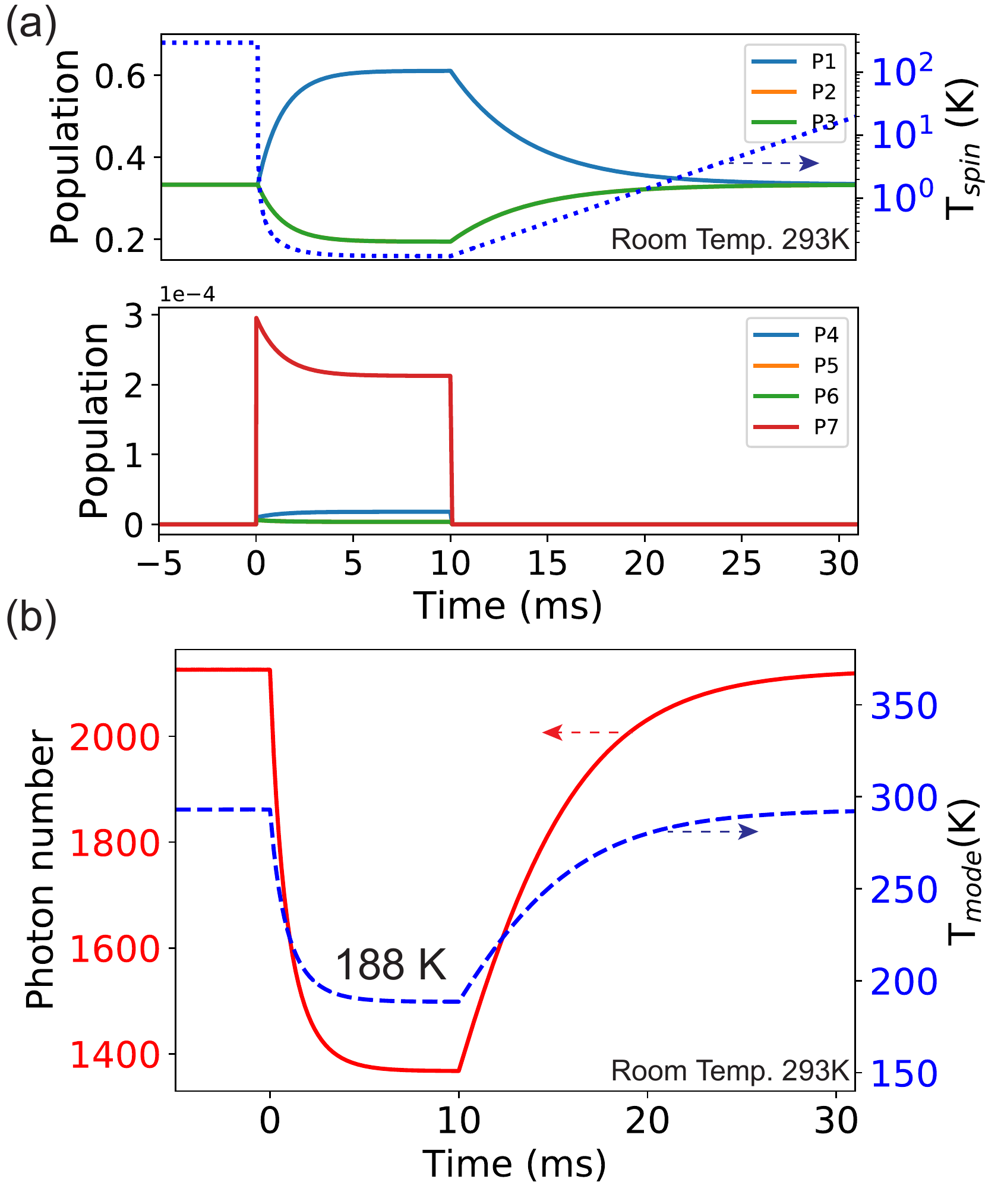}
\par\end{centering}
\caption{\label{fig:coolingdynamics-weak} Similar results as those in Fig. \ref{fig:coolingdynamics-strong} in the main text except for the setup with  lower frequency microwave resonator, as  used in the microwave mode cooling experiment~\citep{NgW}.  }
\end{figure}

\begin{figure}[!htp]
\begin{centering}
\includegraphics[scale=0.55]{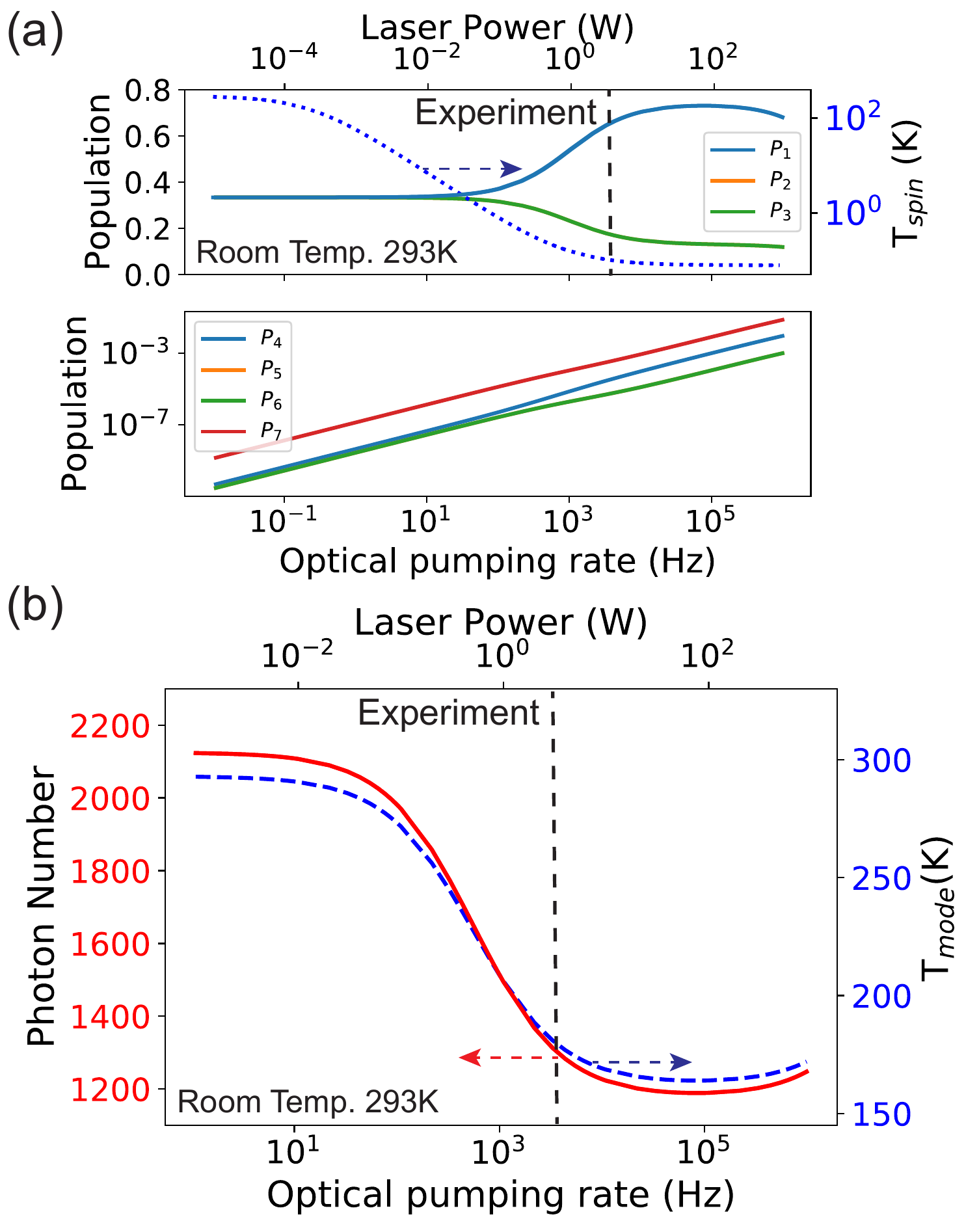}
\par\end{centering}
\caption{\label{fig:ste-state-cooling-weak} Similar results as those in Fig. \ref{fig:ste-state-cooling-weak} in the main text except for the setup with  lower frequency microwave resonator, as  used in the microwave mode cooling experiment~\citep{NgW}.}
\end{figure}

With the transition rates in the upper part of Tab. \ref{tab:param} and the parameters in Tab. \ref{tab:param-cooling}, we have simulated the cooling dynamics (Fig. \ref{fig:coolingdynamics-weak}) and steady-state (Fig. \ref{fig:ste-state-cooling-weak}) of the NV centers and the coupled microwave resonator. Fig. \ref{fig:coolingdynamics-weak} shows similar results as Fig. \ref{fig:coolingdynamics-strong} in the main text except:  (1) the intra-resonator photon number starts from relatively larger thermal value $2125$ because the resonator energy $\hbar\omega_m$ is much smaller than the thermal energy $k_B T$; (2) the effective mode temperature reduces to $188$ K, about $64$ K higher than $124$ K in Fig. \ref{fig:coolingdynamics-strong}(b), due to the insufficient energy transfer rate from the NV centers to the microwave mode.  Fig. \ref{fig:ste-state-cooling-weak} shows similar results as Fig. \ref{fig:ste-state-cooling-strong} in the main text except:  the minimal mode temperature is $170$ K, about $55$ K higher than the value $115$ K in Fig. \ref{fig:ste-state-cooling-strong}(b). The dynamics of the mode temperature reproduces the experiment result \citep{NgW}, and the mode temperature can be reduced from $188$ K to the minimal temperature $170$ K by increasing the laser power from $2$ W to  $100$ W.

Furthermore, we verify that the rate equations given in the Appendix \ref{subsec:elipop} can reproduce perfectly these results, because the system works in the weak coupling regime, and the spin-spin correlation does not play a role here. Since $k_{eet}$ is in the order of $10^{-10}$ Hz while the other rates in the rate equations is in the order of $10^{2},10^{3}$ Hz, we can ignore the terms proportional to the photon number $\langle\hat{a}^{\dagger}\hat{a}\rangle$ in Eqs. (\ref{eq:11-new}) to (\ref{eq:33-new}). Furthermore, since the parameters related to the $\pm1$ spin levels are similar due to their similar energy, the equations for the populations $\langle\hat{\sigma}_{1}^{22}\rangle$ and $\langle\hat{\sigma}_{1}^{33}\rangle$ are equivalent, leading to $\langle\hat{\sigma}_{1}^{33}\rangle\approx\langle\hat{\sigma}_{1}^{22}\rangle\approx(1-\langle\hat{\sigma}_{1}^{11}\rangle)/2$. In this case, we can approximate Eq. (\ref{eq:11-new}) as
\begin{equation}
 \partial_{t}\langle\hat{\sigma}_{1}^{11}\rangle  \approx -(\tilde{k}_{11}+k_{12} + k_{13} + \tilde{k}_{31}  + k_{31})\langle\hat{\sigma}_{1}^{11}\rangle + \tilde{k}_{31}  + k_{31}, 
\end{equation}
and solve it analytically to obtain 
\begin{equation}
\langle\hat{\sigma}_{1}^{11}\rangle (t) \approx (c_1 - c_0)e^{-(\tilde{k}_{11}+k_{12} + k_{13}+\tilde{k}_{31} + k_{31})t} + c_0.
\end{equation}
For the optical spin cooling period in Fig. \ref{fig:coolingdynamics-weak} (a), the two constants are $c_0\approx \frac{\tilde{k}_{31}+k_{31}}{\tilde{k}_{11}+k_{12}+k_{13}+\tilde{k}_{31}+k_{31}}$ and $c_1 \approx 1/3$. For the thermalization period in Fig. \ref{fig:coolingdynamics-weak}(a), the effective rates are $\tilde{k}_{31}=\tilde{k}_{11}=0$ and 
the constants are $c_0 \approx 1/3$ and $c_1\approx \frac{\tilde{k}_{31}+k_{31}}{\tilde{k}_{11}+k_{12}+k_{13}+\tilde{k}_{31}+k_{31}}$. From this population we can compute the population of the $\pm 1$ spin levels $\langle\hat{\sigma}_{1}^{33}\rangle (t)\approx \langle\hat{\sigma}_{1}^{22}\rangle (t)\approx 1- 2\langle\hat{\sigma}_{1}^{11}\rangle (t)$, as well as  the intra-resonator photon number  $\langle\hat{a}^{\dagger}\hat{a}\rangle (t) \approx \frac{N k_{eet} \langle\hat{\sigma}_{1}^{11}\rangle (t) +\kappa n_m^{th} }{N k_{eet} [\langle\hat{\sigma}_{1}^{11}\rangle (t)-\langle\hat{\sigma}_{1}^{33}\rangle (t)]+\kappa}$ with the energy transfer rate $k_{eet}=  \frac{2g_{31}^2 \chi}{\left(\omega_{m}-\omega_{31}\right)^2 + \chi^2}$, where $\chi = \left(\kappa+k_{12}+k_{13}+k_{31}\right)/2+\xi+\chi_{3}$ is   the total dephasing rate. The photon number can be further simplified as $\langle\hat{a}^{\dagger}\hat{a}\rangle (t) \approx \frac{\kappa n_m^{th} }{N k_{eet} [\langle\hat{\sigma}_{1}^{11}\rangle (t)-\langle\hat{\sigma}_{1}^{33}\rangle (t)]+\kappa}$ for $N k_{eet} \ll \kappa n_m^{th}$. The above expressions indicate that the population of the $0$ spin level evolves exponentially during both the optical cooling and thermalization period, and the intra-resonant photon number behaves oppositely to the population.

For the current setup, we estimate the largest NV spins-microwave mode coupling as  $\sqrt{2J}g\approx\sqrt{0.7\times 1.6\times 10^{15}}\times 0.084 \approx 2\pi \times 0.45$ MHz, and conclude that the system is in the collective weak coupling regime since this coupling is much smaller than the spin dephasing rate $\chi_3=2\pi\times 2.6$ MHz. As a result, with the current setup, we can not realize the C-QED effects at room temperature. 

\begin{figure}[!htp]
\begin{centering}
\includegraphics[scale=0.45]{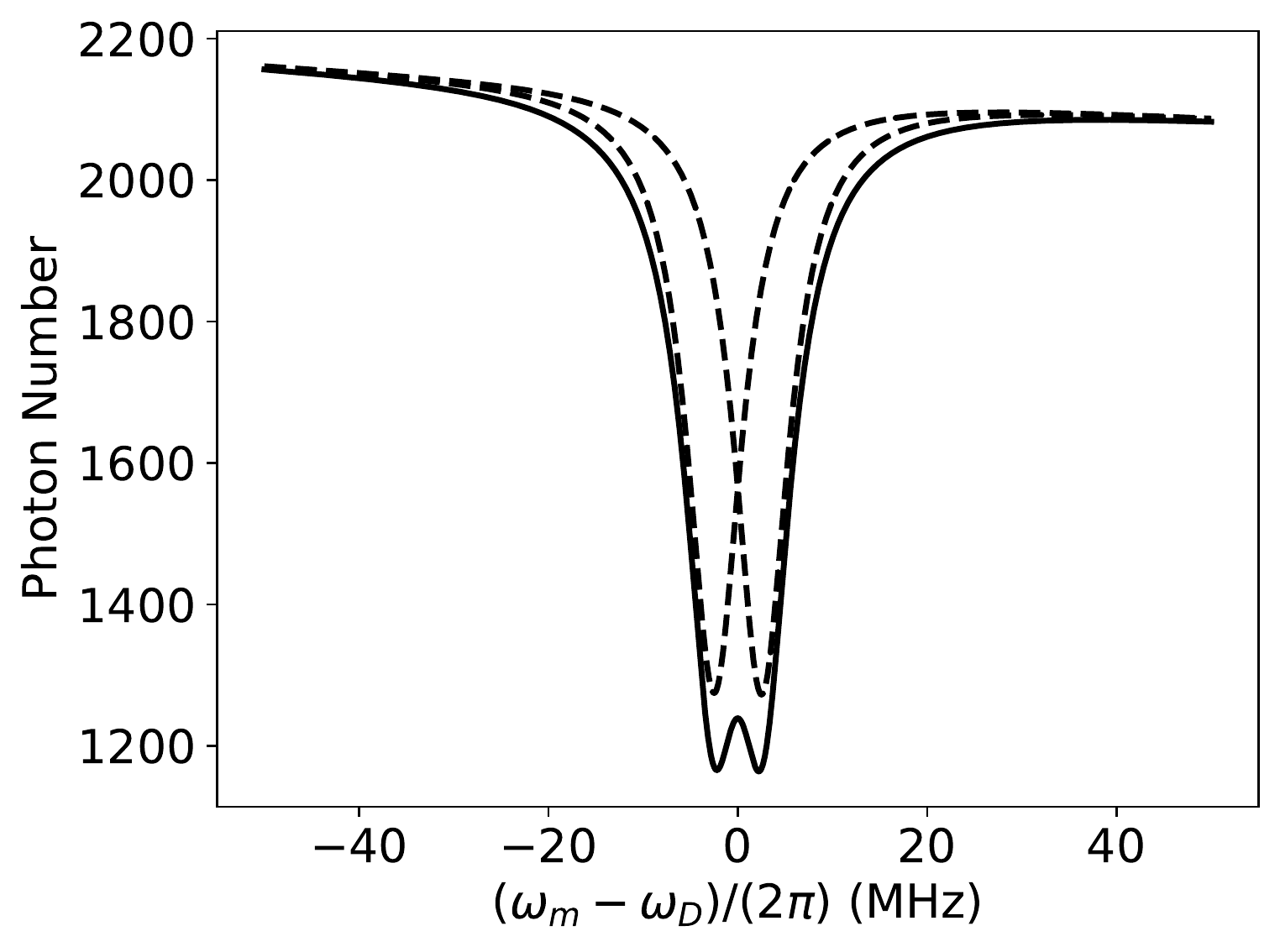}
\par\end{centering}
\caption{\label{fig:phomodshift} Intra-resonator photon number at steady-state as function of the detuning of the microwave mode frequency $\omega_m$ to the middle frequency $\omega_D = (\omega_{31}+\omega_{21})/2$ of the $0\to+1$ and $0\to -1$ transitions, for the laser excitation with the power of $2$ W. The dashed lines indicate the results when only one transition is considered. }
\end{figure}
 
Considering the large spin-dephasing rate, the $-1$ spin level might also influence the microwave mode cooling even though it is off-resonant to the microwave mode. To quantify this influence, we should extend the multi-level JC model in the main text by considering the coupling between this spin level with the microwave mode. However, in the light that this coupling has no influence on the cooling dynamics and steady-state of the NV spin ensemble, we conclude that this coupling affects only Eq. (\ref{eq:apa-analytic}) for the intra-resonator photon number:
\begin{equation}
\langle\hat{a}^{\dagger}\hat{a}\rangle (t) \approx \frac{N (k_{eet}^{31} + k_{eet}^{21}) \langle\hat{\sigma}_{1}^{11}\rangle (t)  +\kappa n_m^{th} }{N \sum_{i=2,3} k_{eet}^{i1} [\langle\hat{\sigma}_{1}^{11}\rangle (t)-\langle\hat{\sigma}_{1}^{ii}\rangle (t)] +\kappa}, \label{eq:apa-analytic-two}
\end{equation}
where we have introduced the energy transfer rates $k_{eet}^{i1}= - 2g_{i1}^2 {\rm Im} \tilde{\delta_{i1}}^{-1}$ with the complex detuning  $\tilde{\delta}_{31} =  \left(\omega_{m}-\omega_{31}\right) + i \left[\left(\kappa+k_{12}+k_{13}+k_{31}\right)/2+\xi+\chi_{3}\right]$ and $\tilde{\delta}_{21} =  \left(\omega_{m}-\omega_{21}\right) + i \left[\left(\kappa+k_{13}+k_{12}+k_{21}\right)/2+\xi+\chi_{2}\right]$. Assuming the same parameters for the $-1\to 0$ and $+1\to 0$ transition, we can use the above expression to compute the intra-resonator photon number at steady-state as function of the microwave mode frequency $\omega_m$ for the laser excitation of power $2$ W, see Fig. \ref{fig:phomodshift}. We see that the two spin transitions contribute together to the microwave mode cooling, and the off-resonant spin transition contributes to the microwave mode cooling by  $5\%$ when the other transition is resonant to the microwave mode. Note that the photon number decreases with increasing microwave mode frequency $\omega_m$ due to the dependence of the thermal photon number $n_m^{th}$ on this frequency.

\section{Rate Equations for Microwave Mode Cooling \label{sec:rateeqn}}

In this Appendix, we explain how to derive the rate equations by ignoring the spin-spin correlation and eliminating the spin-photon correlation, and to further simplify these equations by eliminating the population of triplet and singlet excited states, as well as to obtain the analytical expressions to the steady-state of the NV spins-microwave resonator system. 

\subsection{Eliminating Correlations \label{subsec:elicor}}

To clarify whether the collective coupling plays a role in the microwave mode cooling via the optically cooled NV spin ensemble, we derive the set of rate equations from the mean-field equations given in the previous appendix. To this end, we assume the vanishing spin-spin correlation $\langle\hat{\sigma}_{1}^{31}\hat{\sigma}_{2}^{13}\rangle\approx0$, and then solve Eq. (\ref{eq:correlation}) at steady-state 
\begin{equation}
\langle\hat{a}^{\dagger}\hat{\sigma}_{1}^{13}\rangle \approx g_{31}\tilde{\delta}^{-1}\left[\langle\hat{\sigma}_{1}^{11}\rangle\langle\hat{a}^{\dagger}\hat{a}\rangle - \langle\hat{\sigma}_{1}^{33}\rangle\left(1+\langle\hat{a}^{\dagger}\hat{a}\rangle\right)\right], \label{eq:apsig}
\end{equation}
with the complex frequency detuning $\tilde{\delta} =  \left(\omega_{m}-\omega_{31}\right) + i \left[\left(\kappa+k_{12}+k_{13}+k_{31}\right)/2+\xi+\chi_{3}\right]$. Inserting the above result into Eq. (\ref{eq:photon}), we obtain 
\begin{align}
\partial_{t}\langle\hat{a}^{\dagger}\hat{a}\rangle  & \approx \kappa\left(n_{m}^{th}-\langle\hat{a}^{\dagger}\hat{a}\rangle\right) - Nk_{eet} \langle\hat{\sigma}_{1}^{11}\rangle\langle\hat{a}^{\dagger}\hat{a}\rangle \nonumber \\
 &  + Nk_{eet} \langle\hat{\sigma}_{1}^{33}\rangle\left(1+\langle\hat{a}^{\dagger}\hat{a}\rangle\right), \label{eq:photon-new}
\end{align}
with the energy transfer rate $k_{eet}= - 2g_{31}^2 {\rm Im} \tilde{\delta}^{-1}$ for single NV center to the microwave mode. Inserting Eq. (\ref{eq:apsig}) to Eqs. (\ref{eq:11}) and (\ref{eq:33}), we obtain 
\begin{align}
& \partial_{t}\langle\hat{\sigma}_{1}^{11}\rangle =  - (k_{12}+k_{13} + \xi )\langle\hat{\sigma}_{1}^{11}\rangle + (\xi+k_{sp}) \langle\hat{\sigma}_{1}^{44}\rangle + k_{21}\langle\hat{\sigma}_{1}^{22}\rangle \nonumber \\
 & +k_{31}\langle\hat{\sigma}_{1}^{33}\rangle - k_{eet} \langle\hat{\sigma}_{1}^{11}\rangle\langle\hat{a}^{\dagger}\hat{a}\rangle +k_{eet} \langle\hat{\sigma}_{1}^{33}\rangle\left(1+\langle\hat{a}^{\dagger}\hat{a}\rangle\right), \label{eq:11new} \\
 & \partial_{t}\langle\hat{\sigma}_{1}^{33}\rangle=-\xi\langle\hat{\sigma}_{1}^{33}\rangle+(k_{sp}+\xi)\langle\hat{\sigma}_{1}^{66}\rangle\nonumber \\
 & +k_{73}\langle\hat{\sigma}_{1}^{77}\rangle+k_{13}\langle\hat{\sigma}_{1}^{11}\rangle-k_{31}\langle\hat{\sigma}_{1}^{33}\rangle\nonumber \\
 & + k_{eet} \langle\hat{\sigma}_{1}^{11}\rangle\langle\hat{a}^{\dagger}\hat{a}\rangle -k_{eet} \langle\hat{\sigma}_{1}^{33}\rangle\left(1+\langle\hat{a}^{\dagger}\hat{a}\rangle\right), \label{eq:33new}
\end{align}
Eqs. (\ref{eq:photon-new}), (\ref{eq:11new}) and (\ref{eq:33new}) together with Eqs. (\ref{eq:22}), (\ref{eq:44})-(\ref{eq:77}) constitute the rate equations for the microwave mode cooling. Our equations differ from those in Ref. \citep{NgW} in the following aspects: (1) the energy transfer rate depends on the frequency detuning, and thus our rate equations can be also applied to the off-resonant condition; (2) the energy transfer rate depends also on the resonator mode damping rate $\kappa$ and the optical pumping rate $\xi$, and thus our equations are valid when these two rates become comparable with the spin-dephasing rate $\chi$; (3) our equations account for the resonator-mediated spontaneous emission of microwave photon (i.e. superradiance) and thus are also valid when the microwave mode are cooled to the ground state. However, in the situations as encountered in the experiments~\citep{NgW}, the spin-dephasing rate dominates over other rates and the number of microwave photons is very large. As a result, our rate equations are equivalent to those developed in \citep{NgW}.

\subsection{Eliminating Population of Higher Excited States \label{subsec:elipop}}

In light of Fig. \ref{fig:coolingdynamics-strong} in the main text, the population of the triplet and single excited state seem to follow adiabatically the population on the spin levels of the triplet ground state due to the large spontaneous emission rate compared to other rates. Motivated by this, we might apply the adiabatic approximation to eliminate the population on the triplet and singlet excited states. To this end, we consider the steady-state solutions for these populations from Eqs. (\ref{eq:44})-(\ref{eq:77}): $\langle\hat{\sigma}_{1}^{44}\rangle \approx \langle\hat{\sigma}_{1}^{11} \rangle\frac{\xi}{k_{sp}+\xi+k_{47}}$, $\langle\hat{\sigma}_{1}^{55}\rangle \approx \langle\hat{\sigma}_{1}^{11} \rangle\frac{\xi}{k_{sp}+\xi+k_{57}}$, $\langle\hat{\sigma}_{1}^{66}\rangle \approx \langle\hat{\sigma}_{1}^{11} \rangle\frac{\xi}{k_{sp}+\xi+k_{67}}$,  and $\langle\hat{\sigma}_{1}^{77}\rangle \approx \frac{1}{k_{71}+k_{72}+k_{73}}(k_{47}\langle\hat{\sigma}_{1}^{44}\rangle + k_{57}\langle\hat{\sigma}_{1}^{55}\rangle + k_{67}\langle\hat{\sigma}_{1}^{66}\rangle)$, and insert these results into Eqs. (\ref{eq:22}) and (\ref{eq:33new}) to obtain 
\begin{align}
 &\partial_{t}\langle\hat{\sigma}_{1}^{11}\rangle  =-(\tilde{k}_{11}+k_{12} + k_{13})\langle\hat{\sigma}_{1}^{11}\rangle+(\tilde{k}_{21}+k_{21})\langle\hat{\sigma}_{1}^{22}\rangle  \nonumber  \\
 & + (\tilde{k}_{31}  + k_{31})\langle\hat{\sigma}_{1}^{33}\rangle
  - k_{eet} \langle\hat{\sigma}_{1}^{11}\rangle \langle\hat{a}^{\dagger}\hat{a}\rangle + k_{eet} \langle\hat{\sigma}_{1}^{33}\rangle (1+\langle\hat{a}^{\dagger}\hat{a}\rangle), \label{eq:11-new} \\
&  \partial_{t}\langle\hat{\sigma}_{1}^{22}\rangle=-(\tilde{k}_{22}+k_{21})\langle\hat{\sigma}_{1}^{22}\rangle+(\tilde{k}_{12}+k_{12})\langle\hat{\sigma}_{1}^{11}\rangle + \tilde{k}_{31}\langle\hat{\sigma}_{1}^{33}\rangle, \label{eq:22-new} \\
 &\partial_{t}\langle\hat{\sigma}_{1}^{33}\rangle  =-(\tilde{k}_{33}+k_{31})\langle\hat{\sigma}_{1}^{22}\rangle+(\tilde{k}_{13}+k_{13})\langle\hat{\sigma}_{1}^{11}\rangle \nonumber \\ 
 &+ \tilde{k}_{23}\langle\hat{\sigma}_{1}^{22}\rangle  + k_{eet} \langle\hat{\sigma}_{1}^{11}\rangle \langle\hat{a}^{\dagger}\hat{a}\rangle - k_{eet} \langle\hat{\sigma}_{1}^{33}\rangle (1+\langle\hat{a}^{\dagger}\hat{a}\rangle). \label{eq:33-new}
\end{align}
In the above equations, we have defined the rates $\tilde{k}_{11} = \xi \frac{k_{47}}{k_{sp}+\xi +k_{47}}(1-\frac{k_{71}}{k_{71}+k_{72}+k_{73}})$, $\tilde{k}_{21} = \xi \frac{k_{57}}{k_{sp}+\xi +k_{57}}\frac{k_{71}}{k_{71}+k_{72}+k_{73}}$, $\tilde{k}_{31} = \xi \frac{k_{67}}{k_{sp}+\xi +k_{67}}\frac{k_{71}}{k_{71}+k_{72}+k_{73}}$, and $\tilde{k}_{12} = \xi \frac{k_{47}}{k_{sp}+\xi +k_{47}}\frac{k_{72}}{k_{71}+k_{72}+k_{73}}$, $\tilde{k}_{22} = \xi \frac{k_{57}}{k_{sp}+\xi +k_{57}}(1-\frac{k_{72}}{k_{71}+k_{72}+k_{73}})$, $\tilde{k}_{32} = \xi \frac{k_{67}}{k_{sp}+\xi +k_{67}}\frac{k_{72}}{k_{71}+k_{72}+k_{73}}$, as well as $\tilde{k}_{13} = \xi \frac{k_{47}}{k_{sp}+\xi +k_{47}}\frac{k_{73}}{k_{71}+k_{72}+k_{73}}$, $\tilde{k}_{23} = \xi \frac{k_{57}}{k_{sp}+\xi +k_{57}}\frac{k_{73}}{k_{71}+k_{72}+k_{73}}$, $\tilde{k}_{33} = \xi \frac{k_{67}}{k_{sp}+\xi +k_{67}}(1-\frac{k_{73}}{k_{71}+k_{72}+k_{73}})$. Here, $\tilde{k}_{ii}$  are the rates of the population transfer from the spin levels of the triplet ground state to the higher excited states, where $\tilde{k}_{ij}$ with $i \neq j=1,2,3$ are the rates of the population transfer from the $i$-th spin level to the $j$-th spin level of the triplet ground state via the higher excited states. 

Our calculations show that $k_{eet}$ is in the order of $10^{-7}$ Hz while the other rates in the above equations is in the order of $10^{2},10^{3}$ Hz. Thus, in Eqs. (\ref{eq:11-new}) to (\ref{eq:33-new}), we can ignore the terms proportional to the photon number $\langle\hat{a}^{\dagger}\hat{a}\rangle$. Furthermore, the photon damping rate $\kappa$ and the collective energy transfer rate $N k_{eet}$ are in the order of $10^{6}$ Hz, which is orders of magnitude larger than the rates appearing in  Eqs. (\ref{eq:11-new}) to (\ref{eq:33-new}) for the spin level populations. As a result, we might also apply the adiabatic condition to Eq. (\ref{eq:photon-new}) to obtain 
\begin{equation}
\langle\hat{a}^{\dagger}\hat{a}\rangle (t) \approx \frac{N k_{eet} \langle\hat{\sigma}_{1}^{11}\rangle (t) +\kappa n_m^{th} }{N k_{eet} [\langle\hat{\sigma}_{1}^{11}\rangle (t)-\langle\hat{\sigma}_{1}^{33}\rangle (t)]+\kappa}. \label{eq:apa-analytic}
\end{equation}

\subsection{Analytical Expressions \label{subsec:analytic}}

In this part, we consider the steady-state situation and derive the corresponding analytical expressions. To proceed, we ignore the coupling with the microwave mode in Eqs. (\ref{eq:22-new}) and (\ref{eq:33-new}), and consider the equations at steady-state. Furthermore, using the relations $\langle\hat{\sigma}_{1}^{11}\rangle=1-\langle\hat{\sigma}_{1}^{22}\rangle - \langle\hat{\sigma}_{1}^{33}\rangle$, we can solve these equations analytically and obtain 
$\langle\hat{\sigma}_{1}^{22}\rangle	\approx \frac{B\left(\tilde{k}_{12}+k_{12}\right)-C\left(\tilde{k}_{13}+k_{13}\right)}{AB-CD},$
$\langle\hat{\sigma}_{1}^{33}\rangle	\approx \frac{A\left(\tilde{k}_{13}+k_{13}\right)-D\left(\tilde{k}_{12}+k_{12}\right)}{AB-CD}$ with the abbreviations $A=\tilde{k}_{22}+k_{21}+\tilde{k}_{12}+k_{12}, B= \tilde{k}_{33}+k_{31}+\tilde{k}_{13}+k_{13}, C = \tilde{k}_{12}+k_{12}-\tilde{k}_{32}, D=  \tilde{k}_{13}+k_{13}-\tilde{k}_{23}$. We have verified that these analytical expressions can reproduce the results obtained with the rate equations.

\section{ Counteracting Optical Heating of Diamond \label{sec:opticalheating}}

The  results in the main text show that the laser power about $1$ W is enough to achieve the microwave mode cooling and the C-QED effects at room temperature, and the laser power about $100$ W is necessary to achieve the optimized spin cooling, and thus more better performance. Under such strong laser illumination, the diamond might be overheated. It was reported in the experiment of the diamond maser at room temperature~\citep{Breeze} that the temperature increases by $35$ K when the laser power increases to $0.4$ W, leading to the relationship $\Delta T/\Delta P = 87.5$ ${\rm K/W}$.  According to Ref.~\citep{AcostaVM}, the change of the zero-field splitting depends on the temperature change as $\Delta D/\Delta T = -2\pi \times 0.074 {\rm MHz/K}$. 

Assuming the similar property of the diamond sample, we expect that for the laser power $P=1$ W, the temperature raises by $87.5$ K, and the frequency is drifted by $-2\pi\times6.48$ MHz. This simple calculation indicates that we do need to migrate the optical heating effect. To counteract the optical heating, we can cool the diamond with heat-sinking, forced air, immersive liquid cooling or heat pipes, as pointed out in Ref.~\citep{NgW}. As an example, in the experiment~\citep{AdrishaS}, it was demonstrated that by cooling the diamond sample with water and $N_2$ gas, the temperature raise can be managed within $30$ K for the laser power up to $24$ W. If this cooling is still efficient for much larger laser power, we expect the temperature rise of $125$ K for the laser power $100$ W, and the drift of the frequency by $2\pi\times 9.25$ MHz. This kind of heating is still accepted and the frequency drift can be counteracted by tunning the Zeeman shift induced by a proper magnetic field. 

The temperature discussed above refers to that of the diamond lattice phonon $T_{lat}$ not the NV spins and the microwave mode, and affects the phenomena studied in the main text only by increasing spin-lattice relaxation rate. According to Refs. ~\citep{Anorambuena,AJarmola}, for the lattice temperature above $200$ K , the spin-lattice relaxation rate is determined by the two-phonon Raman process, and scales as $k_{i1} \propto T_{lat}^5$ with the lattice temperature. As a result, we have $k_{i1}\approx k_{1i}$ ($i=2,3$) to $k_{i1}\times{[(T_{ini}+\Delta T)/T_{ini}]^5}$ with $T_{ini}=293$ K and $\Delta T=125$ K. For the setup with high frequency resonator, $k_{31}$ increases from $26$ Hz to $154$ Hz, and $k_{31}$ increases from $98$ Hz to $140$ Hz. For the setup with lower frequency resonator, $k_{31}\approx k_{21}$ increases from $83$ Hz to $490$ Hz. For the increased spin-lattice relaxation rate, we can still obtain the results as in the main text except that the laser power should be increased slightly to compensate the enhanced spin-lattice relaxation.


\end{document}